\documentclass[iop,apj]{emulateapj}
\usepackage{apjfonts}
\usepackage[normalem]{ulem}

\journalinfo{The Astrophysical Journal}
\slugcomment{accepted for publication, July 15, 2015}

\shortauthors{CAMILO ET AL.}
\shorttitle{PARKES PULSAR SEARCHES OF FERMI SOURCES}

\begin{document}

\def\fermi{{\em Fermi}}
\def\swift{{\em Swift}}
\def\cxo{{\em CXO}}
\def\hr{^{\rm h}}
\def\mi{^{\rm m}}
\def\s{^{\rm s}}
\def\msun{M$_{\sun}$}

\def\psra{PSR~J1514$-$4946}
\def\psrb{PSR~J1658$-$5324}
\def\psrc{PSR~J1747$-$4036}
\def\psrd{PSR~J1902$-$5105}
\def\psre{PSR~J0101$-$6422}
\def\psrf{PSR~J0955$-$6150}
\def\psrg{PSR~J1012$-$4235}
\def\psrh{PSR~J1036$-$8317}
\def\psri{PSR~J1903$-$7051}
\def\psrj{PSR~J1946$-$5403}

\title{Parkes radio searches of {\em Fermi\/} gamma-ray sources and
millisecond pulsar discoveries }

\author{F.~Camilo\altaffilmark{1},
  M.~Kerr\altaffilmark{2},
  P.~S.~Ray\altaffilmark{3},
  S.~M.~Ransom\altaffilmark{4},
  J.~Sarkissian\altaffilmark{5},
  H.~T.~Cromartie\altaffilmark{6},
  S.~Johnston\altaffilmark{2},
  J.~E.~Reynolds\altaffilmark{2},
  M.~T.~Wolff\altaffilmark{3},
  P.~C.~C.~Freire\altaffilmark{7},
  B.~Bhattacharyya\altaffilmark{8},
  E.~C.~Ferrara\altaffilmark{9},
  M.~Keith\altaffilmark{8},
  P.~F.~Michelson\altaffilmark{10},
  P.~M.~Saz~Parkinson\altaffilmark{11,12},
  and K.~S.~Wood\altaffilmark{3}
}

\altaffiltext{1}{Columbia Astrophysics Laboratory, Columbia University,
  New York, NY~10027, USA}
\altaffiltext{2}{CSIRO Astronomy and Space Science, Australia
  Telescope National Facility, Epping, NSW~1710, Australia}
\altaffiltext{3}{Space Science Division, Naval Research Laboratory,
  Washington, DC~20375-5352, USA}
\altaffiltext{4}{National Radio Astronomy Observatory, Charlottesville,
  VA~22903, USA}
\altaffiltext{5}{CSIRO Parkes Observatory, Parkes, NSW~2870, Australia}
\altaffiltext{6}{Department of Astronomy, University of Virginia,
  Charlottesville, VA~22904-4325, USA}
\altaffiltext{7}{Max-Planck-Institut f\"{u}r Radioastronomie,
  D-53121 Bonn, Germany}
\altaffiltext{8}{Jodrell Bank Centre for Astrophysics, School of
  Physics and Astronomy, The University of Manchester, Manchester M13
  9PL, UK}
\altaffiltext{9}{NASA Goddard Space Flight Center, Greenbelt, MD 20771, USA}
\altaffiltext{10}{W.~W.~Hansen Experimental Physics Laboratory,
  Kavli Institute for Particle Astrophysics and Cosmology, Department
  of Physics and SLAC National Accelerator Laboratory, Stanford
  University, Stanford, CA 94305, USA}
\altaffiltext{11}{Santa Cruz Institute for Particle Physics,
  Department of Physics and Department of Astronomy and Astrophysics,
  University of California at Santa Cruz, Santa Cruz, CA 95064, USA}
\altaffiltext{12}{Department of Physics, The University of Hong
  Kong, Pokfulam Road, Hong Kong, China}

\begin{abstract}
In a search with the Parkes radio telescope of 56 unidentified
\fermi-LAT gamma-ray sources, we have detected 11 millisecond pulsars
(MSPs), 10 of them discoveries, of which five were reported in
\citet{kcj+12}.  We did not detect radio pulsations from another
six pulsars now known in these sources.  We describe the completed
survey, which included multiple observations of many targets done
to minimize the impact of interstellar scintillation, acceleration
effects in binary systems, and eclipses.  We consider that 23 of
the 39 remaining sources may still be viable pulsar candidates.  We
present timing solutions and polarimetry for five of the MSPs, and
gamma-ray pulsations for \psri\ (pulsations for five others were
reported in the second \fermi-LAT catalog of gamma-ray pulsars).
Two of the new MSPs are isolated and five are in $>1$\,d circular
orbits with 0.2--0.3\,\msun\ presumed white dwarf companions. \psrf,
in a 24\,d orbit with a $\approx 0.25$\,\msun\ companion but
eccentricity of 0.11, belongs to a recently identified class of
eccentric MSPs.  \psrh\ is in an 8\,hr binary with a $>0.14$\,\msun\
companion that is probably a white dwarf.  \psrj\ is in a 3\,hr
orbit with a $>0.02$\,\msun\ companion with no evidence of radio
eclipses.

\end{abstract}


\section{Introduction} \label{sec:intro} 

The Large Area Telescope \citep[LAT;][]{aaa+09m} on the {\em Fermi
Gamma-ray Space Telescope\/} is a superb instrument with which to
study rotation-powered pulsars.  Since mid-2008, it has been used
to detect more than 170 pulsars at energies above 0.1\,GeV
\citep[e.g.,][]{aaa+10c}\footnote{https://confluence.slac.stanford.edu/display/GLAMCOG/Public+List+of+LAT-Detected+Gamma-Ray+Pulsars.}.
\fermi\ has identified millisecond pulsars (MSPs) as a ubiquitous
class of gamma-ray sources \cite[e.g.,][]{aaa+09f}.  These old
neutron stars, spun up by accretion from an evolved companion
\citep{acrs82}, are a relatively local and isotropically distributed
population, that make up only 10\% of the identified population of
pulsars in the Galactic disk.  However, about half of known gamma-ray
pulsars are MSPs.

Nearly half of the known gamma-ray MSPs were discovered as radio
objects in undirected (``all sky'') surveys prior to the launch of
\fermi.  Gamma-ray pulsations were subsequently detected with the
aid of rotational ephemerides obtained from radio timing observations.
Many slowly-rotating pulsars have been discovered via direct
periodicity searches of sparse gamma-ray photons
\citep[e.g.,][]{aaa+09c,pga+12}, but so far this has not been
possible unbiasedly for binary MSPs.  However, the LAT has led to
the discovery of MSPs in a different, and prolific, fashion.  In
the three LAT source catalogs \citep{aaa+10g,naa+12,aaa+15}, there
are hundreds of unidentified sources, many of which have spectral
characteristics typical of pulsars.  Radio searches of many of these
have turned up dozens of MSPs so far, and once the radio ephemerides
have been obtained, gamma-ray pulsations have almost always followed
\citep[e.g.,][]{cgj+11,bgc+13}.

Using the CSIRO Parkes telescope in 2009, we discovered five MSPs
in a radio survey of 14 unidentified LAT sources \citep{kcj+12}.
In an extension of that survey we have discovered five more MSPs.
Here we present the completed survey, and report on radio timing,
polarimetric, and gamma-ray studies of some of the MSPs.

\section{Observations, Analysis, and Results} \label{sec:obs} 

In this Section we describe the radio searches of unidentified
gamma-ray sources that we performed at the Parkes telescope
(Section~\ref{sec:searches}), the sensitivity of the survey and
relevant selection effects (Section~\ref{sec:sensitivity}), radio
timing and polarimetric observations of the pulsars discovered
(Sections~\ref{sec:timing} and \ref{sec:pol}), gamma-ray results
for one of them (Section~\ref{sec:gamma}), and some X-ray observations
(Section~\ref{sec:xray}).

\subsection{Parkes Radio Searches} \label{sec:searches}

\subsubsection{Initial Searches of 1FGL Sources} \label{sec:1fgl}

\citet{kjr+11} used a digital filterbank at Parkes to search 11
unidentified sources from the first \fermi\ LAT catalog
\citep[1FGL;][]{aaa+10g}. Two MSPs and one slow pulsar were discovered
in single observations of each target at a central frequency of
1.4\,GHz. Subsequently, one of these MSPs was found to be associated
with the corresponding LAT source.

At nearly the same time, in late 2009, we used an analog filterbank
at Parkes to search 14 unidentified 1FGL sources \citep{kcj+12}.
The single observations of these targets resulted in the detection
of six MSPs, five of them discoveries. However, confirmation of
some of these MSPs was not easy: the search observations lasted for
1--2\,hr each, but some of the pulsars were not detected in equivalent
initial confirmation attempts, owing to the effects of interstellar
scintillation. These and other selection effects (see
Section~\ref{sec:sensitivity} for details) led us to search some
promising unidentified LAT sources repeatedly.

\subsubsection{Repeated Searches of Unidentified LAT Sources} \label{sec:2fgl}

Our subsequent searches used the same equipment and methods as
\citet{kcj+12}.  In brief, total-power measures from the central
beam of the 20\,cm multibeam receiver were filtered into 512
contiguous 0.5\,MHz-wide channels centered on 1390\,MHz and sampled
8000 times per second, then digitized with 1-bit precision and
written to disk for off-line analysis.  Individual integration times
were about 1\,hr, and each LAT source was observed between one and
nine times, depending on the then-perceived quality of the source
and telescope availability (Table~\ref{tab:survey}).

\begin{deluxetable*}{lccrrcr}
\tabletypesize{\scriptsize}
\tablewidth{0.99\linewidth}
\tablecaption{\label{tab:survey} Radio Searches of FGL Sources at Parkes: Observations }
\tablecolumns{7}
\tablehead{
\colhead{Name\tablenotemark{a}}                  &
\colhead{R.A.\tablenotemark{b}}                  &
\colhead{Decl.\tablenotemark{b}}                 &
\colhead{$l$}                                    &
\colhead{$b$}                                    &
\colhead{Integration time}                       &
\colhead{$\mbox{DM}_{\rm max}$\tablenotemark{c}} \\
\colhead{}                &
\colhead{(J2000.0)}       &
\colhead{(J2000.0)}       &
\colhead{(deg)}           &
\colhead{(deg)}           &
\colhead{(min)}           &
\colhead{(pc\,cm$^{-3}$)}
}
\startdata
{\bf 3FGL~J0101.0$-$6422} & $01\hr00\mi58\s$ & $-64\arcdeg24'06\arcsec$ & 301.2 & $-$52.7 & {\bf 60} & 270 \\
3FGL~J0602.8$-$4016 & $06\hr03\mi05\s$ & $-40\arcdeg11'04\arcsec$ & 246.8 & $-$25.9 & 90, 120 & 270 \\
{\em 3FGL~J0933.9$-$6232} & $09\hr34\mi00\s$ & $-62\arcdeg30'17\arcsec$ & 282.2 &  $-$7.8 & 120, 60, 60, 60, 60 & 435 \\
{\em 3FGL~J1035.7$-$6720} & $10\hr36\mi10\s$ & $-67\arcdeg21'05\arcsec$ & 290.4 &  $-$7.8 & 120, 86, 136, 60, 60, 60 & 435 \\
{\em 3FGL~J1227.9$-$4854} & $12\hr27\mi50\s$ & $-48\arcdeg51'57\arcsec$ & 298.9 &   13.8 & 120, 65, 60 & 307\\
{\em 3FGL~J1231.6$-$5113} & $12\hr31\mi49\s$ & $-51\arcdeg18'49\arcsec$ & 299.8 &   11.4 & 120 & 270 \\
{\bf 3FGL~J1514.2$-$4947} & $15\hr14\mi06\s$ & $-49\arcdeg45'33\arcsec$ & 325.2 &    6.8 & {\bf 120} & 270 \\
{\em 3FGL~J1624.2$-$4041} & $16\hr24\mi07\s$ & $-40\arcdeg41'20\arcsec$ & 340.6 &    6.2 & 120, 120, 120, 60, 60, 60 & 435 \\
{\bf 3FGL~J1658.4$-$5323} & $16\hr58\mi43\s$ & $-53\arcdeg17'43\arcsec$ & 335.0 &  $-$6.6 & {\bf 83} & 270 \\
{\em 3FGL~J1744.1$-$7619} & $17\hr44\mi02\s$ & $-76\arcdeg20'25\arcsec$ & 317.1 & $-$22.5 & 41, 90, 80, 72, 71, 78, 60, 60, 60 & 192\\
{\bf 3FGL~J1747.6$-$4037} & $17\hr47\mi29\s$ & $-40\arcdeg36'07\arcsec$ & 350.2 &  $-$6.4 & {\bf 86} & 270 \\
{\bf 3FGL~J1902.0$-$5107} & $19\hr02\mi05\s$ & $-51\arcdeg09'45\arcsec$ & 345.6 & $-$22.4 & {\bf 70} & 270 \\
{\em 3FGL~J2039.6$-$5618} & $20\hr39\mi30\s$ & $-56\arcdeg20'45\arcsec$ & 341.2 & $-$37.1 & 75, 112, 60 & 270 \\
{\bf 3FGL~J2241.6$-$5237}\tablenotemark{d} & $22\hr41\mi52\s$ & $-52\arcdeg37'38\arcsec$ & 337.4 & $-$54.9 & {\bf 120} & 270 \\
\tableline
3FGL~J0133.0$-$4413 & $01\hr33\mi27\s$ & $-44\arcdeg08'29\arcsec$ & 279.2 & $-$71.0 & 60, 60, 60, 60, 60, 60 &  77 \\
3FGL~J0216.1$-$7016 & $02\hr14\mi04\s$ & $-69\arcdeg52'06\arcsec$ & 292.9 & $-$45.6 & 60 & 115 \\
3FGL~J0744.8$-$4028 & $07\hr44\mi59\s$ & $-40\arcdeg29'49\arcsec$ & 254.6 &  $-$8.0 & 60 & 717 \\
3FGL~J0802.3$-$5610 & $08\hr02\mi46\s$ & $-56\arcdeg15'32\arcsec$ & 270.0 & $-$13.2 & 60, 60, 35, 35, 35 & 627 \\
3FGL~J0933.9$-$6232 & $09\hr34\mi02\s$ & $-62\arcdeg31'34\arcsec$ & 282.2 &  $-$7.9 & 60, 35 & 627 \\
3FGL~J0940.6$-$7609 & $09\hr40\mi45\s$ & $-76\arcdeg09'34\arcsec$ & 292.2 & $-$17.4 & 60 & 269 \\
3FGL~J0940.7$-$6102 & $09\hr40\mi57\s$ & $-61\arcdeg05'10\arcsec$ & 281.9 &  $-$6.3 & 60 & 653 \\
3FGL~J0954.8$-$3948 & $09\hr55\mi03\s$ & $-39\arcdeg49'25\arcsec$ & 269.9 &    11.5 & 60, 60, 60, 60, 60 & 371 \\
{\bf 3FGL~J0955.6$-$6148} & $09\hr55\mi39\s$ & $-61\arcdeg48'36\arcsec$ & 283.7 &  $-$5.7 & {\bf 60}, {\bf 60} & 653 \\
{\bf 3FGL~J1012.0$-$4235} & $10\hr12\mi07\s$ & $-42\arcdeg35'01\arcsec$ & 274.2 &   11.2 & {\bf 60} & 371 \\
3FGL~J1025.1$-$6507 & $10\hr25\mi00\s$ & $-65\arcdeg07'22\arcsec$ & 288.3 &  $-$6.5 & 60 & 755 \\
3FGL~J1035.7$-$6720 & $10\hr36\mi09\s$ & $-67\arcdeg22'17\arcsec$ & 290.4 &  $-$7.8 & 45, 60, 60 & 563 \\
{\bf 3FGL~J1036.0$-$8317} & $10\hr36\mi20\s$ & $-83\arcdeg17'01\arcsec$ & 298.9 & $-$21.5 & 60, 60, {\bf 60} & 192 \\
3FGL~J1057.7$-$6624 & $10\hr58\mi43\s$ & $-66\arcdeg21'59\arcsec$ & 292.0 &  $-$5.9 & 60, 60, 60, 60 & 755 \\
3FGL~J1136.6$-$6826 & $11\hr36\mi47\s$ & $-68\arcdeg25'37\arcsec$ & 296.1 &  $-$6.5 & 60 & 755 \\
3FGL~J1227.9$-$4854 & $12\hr27\mi42\s$ & $-48\arcdeg53'28\arcsec$ & 298.9 &    13.8 & 60, 60 & 371 \\
3FGL~J1231.6$-$5113 & $12\hr31\mi34\s$ & $-51\arcdeg12'31\arcsec$ & 299.8 &    11.5 & 60 & 435 \\
3FGL~J1238.3$-$4543 & $12\hr38\mi19\s$ & $-45\arcdeg43'18\arcsec$ & 300.5 &    17.1 & 60 & 269 \\
3FGL~J1306.8$-$4031 & $13\hr06\mi52\s$ & $-40\arcdeg32'35\arcsec$ & 306.1 &    22.2 & 60 & 192 \\
3FGL~J1311.8$-$3430 & $13\hr11\mi46\s$ & $-34\arcdeg29'19\arcsec$ & 307.7 &    28.2 & 60, 58, 60, 58 & 154 \\
3FGL~J1325.2$-$5411 & $13\hr25\mi15\s$ & $-54\arcdeg11'13\arcsec$ & 307.9 &     8.4 & 60 & 627 \\
3FGL~J1326.7$-$4727 & $13\hr26\mi40\s$ & $-47\arcdeg27'52\arcsec$ & 309.1 &    15.0 & 60, 60 & 371 \\
3FGL~J1417.5$-$4402 & $14\hr17\mi30\s$ & $-44\arcdeg02'40\arcsec$ & 318.9 &    16.1 & 60 & 307 \\
3FGL~J1417.7$-$5026 & $14\hr17\mi39\s$ & $-50\arcdeg25'33\arcsec$ & 316.7 &    10.1 & 60 & 627 \\
3FGL~J1518.2$-$5232 & $15\hr18\mi27\s$ & $-52\arcdeg33'57\arcsec$ & 324.3 &     4.1 & 60, 60, 60, 60 & 1044 \\
3FGL~J1536.3$-$4949 & $15\hr36\mi29\s$ & $-49\arcdeg49'45\arcsec$ & 328.2 &     4.8 & 37, 37, 16, 60, 60, 60 & 883 \\
3FGL~J1539.2$-$3324 & $15\hr39\mi15\s$ & $-33\arcdeg25'42\arcsec$ & 338.7 &    17.5 & 60, 15, 52, 60 & 269 \\
3FGL~J1603.7$-$6011 & $16\hr03\mi44\s$ & $-60\arcdeg11'12\arcsec$ & 324.8 &  $-$5.7 & 60 & 883 \\
3FGL~J1617.4$-$5846 & $16\hr17\mi28\s$ & $-58\arcdeg46'19\arcsec$ & 327.0 &  $-$5.9 & 60 & 883 \\
3FGL~J1624.2$-$4041 & $16\hr24\mi09\s$ & $-40\arcdeg40'23\arcsec$ & 340.6 &     6.2 & 60, 60 & 883 \\
3FGL~J1702.8$-$5656 & $17\hr02\mi33\s$ & $-56\arcdeg54'54\arcsec$ & 332.4 &  $-$9.2 & 60, 60, 60, 60 & 627 \\
3FGL~J1717.4$-$5157 & $17\hr17\mi35\s$ & $-51\arcdeg57'58\arcsec$ & 337.7 &  $-$8.1 & 60 & 755 \\
3FGL~J1736.2$-$4444 & $17\hr36\mi13\s$ & $-44\arcdeg44'50\arcsec$ & 345.5 &  $-$6.7 & 60 & 883 \\
3FGL~J1744.1$-$7619 & $17\hr44\mi11\s$ & $-76\arcdeg20'29\arcsec$ & 317.1 & $-$22.5 & 60 & 192 \\
3FGL~J1753.6$-$4447 & $17\hr53\mi38\s$ & $-44\arcdeg47'50\arcsec$ & 347.1 &  $-$9.4 & 60 & 627 \\
3FGL~J1803.3$-$6706 & $18\hr03\mi25\s$ & $-67\arcdeg07'14\arcsec$ & 326.9 & $-$20.4 & 60 & 230 \\
3FGL~J1808.3$-$3357 & $18\hr08\mi22\s$ & $-33\arcdeg56'05\arcsec$ & 358.1 &  $-$6.7 & 56, 60, 60, 60 & 755 \\
3FGL~J1831.6$-$6503 & $18\hr31\mi45\s$ & $-65\arcdeg05'19\arcsec$ & 329.9 & $-$22.4 & 60 & 192 \\
{\bf 3FGL~J1903.6$-$7052} & $19\hr02\mi43\s$ & $-70\arcdeg53'49\arcsec$ & 324.3 & $-$26.4 & {\bf 60} & 269 \\
{\bf 3FGL~J1946.4$-$5403} & $19\hr46\mi24\s$ & $-54\arcdeg02'46\arcsec$ & 343.9 & $-$29.6 & 60, {\bf 60}, {\bf 60}, 60, {\bf 60}, 54 & 154 \\
3FGL~J1959.8$-$4725 & $19\hr59\mi57\s$ & $-47\arcdeg26'32\arcsec$ & 351.8 & $-$30.9 & 58, 60, 60, 35, 60 & 307 \\
3FGL~J2039.6$-$5618 & $20\hr39\mi51\s$ & $-56\arcdeg20'26\arcsec$ & 341.2 & $-$37.2 & 60, 60, 60, 60 & 115 \\
3FGL~J2043.8$-$4801 & $20\hr43\mi49\s$ & $-48\arcdeg00'45\arcsec$ & 351.7 & $-$38.3 & 60, 60 & 115 \\
3FGL~J2112.5$-$3044 & $21\hr12\mi36\s$ & $-30\arcdeg42'37\arcsec$ &  14.9 & $-$42.4 & 56, 60, 60, 60, 60, 60 & 103 \\
3FGL~J2131.1$-$6625 & $21\hr31\mi06\s$ & $-66\arcdeg24'42\arcsec$ & 326.7 & $-$40.3 & 60, 60 & 115 \\
3FGL~J2133.0$-$6433 & $21\hr33\mi30\s$ & $-64\arcdeg31'58\arcsec$ & 328.7 & $-$41.3 & 60, 60, 60, 60, 60, 60, 60 & 103 \\
3FGL~J2200.0$-$6930 & $22\hr00\mi00\s$ & $-69\arcdeg31'37\arcsec$ & 321.3 & $-$40.9 & 60 & 115 \\
3FGL~J2220.6$-$6833 & $22\hr20\mi24\s$ & $-68\arcdeg32'22\arcsec$ & 320.8 & $-$43.0 & 60, 60, 60 & 115 \\
3FGL~J2333.0$-$5525 & $23\hr33\mi04\s$ & $-55\arcdeg25'32\arcsec$ & 324.2 & $-$58.3 & 60, 60 &  87
\enddata
\tablecomments{Boldfaced entries denote observations with detection
of MSPs.  Discoveries in single-observation searches of the first
14 entries (listed above the horizontal line) were reported in
\citet{kcj+12}.  Italicized entries denote sources re-observed in
2012 at the improved locations listed below the horizontal line.}
\tablenotetext{a}{The names given are of the 3FGL sources closest
to our pointing locations (the offset between the pointing positions
given here and the 3FGL positions are provided in Table~\ref{tab:survey2}). }
\tablenotetext{b}{Parkes telescope pointing position. }
\tablenotetext{c}{Maximum trial dispersion measure used in our
analysis of the respective data set(s), corresponding approximately
to twice the maximum DM predicted by the \citet{cl02} model for the
corresponding line of sight. The first observation of each source
above the horizontal line was analyzed with $\mbox{DM}_{\rm max}
= 270$\,pc\,cm$^{-3}$ \citep{kcj+12}. }
\tablenotetext{d}{MSP discovered independently by \citet{kjr+11}. }
\end{deluxetable*}

All the data were analyzed using PRESTO \citep{ran01}, which
implements standard pulsar search techniques including radio-frequency
interference excision and optimization for signals with changing
apparent spin periods caused by orbital motion.  Finite sampling
time and smearing from dispersive propagation delays within
finite-width filterbank channels both unavoidably degrade sensitivity
to pulsed signals.  Additional smearing can result from use of an
incorrect dispersion measure (DM) to remove the delays between
channels. We dedispersed each observation using a set of trial DMs
such that this effect was negligible.  This was done up to twice
the maximum DM predicted by the \citet{cl02} model for the corresponding
line of sight (see Table~\ref{tab:survey}).  The maximum acceleration
searched for corresponded to signals drifting by $\pm 200/n_h$ bins
in the Fourier domain, where $n_h$ is the largest harmonic at which
a signal is detected (up to 16 harmonics were summed, in powers of
two). This was parameterized by ${\tt zmax}=200$ within PRESTO
\citep{rem02}.

In the 2009 searches (Section~\ref{sec:1fgl}) we detected six MSPs
in single observations of 14 LAT sources (Table~\ref{tab:survey},
above the horizontal dividing line). In 2010--2011 we re-observed
seven of the remaining eight 1FGL sources (Table~\ref{tab:survey},
above the dividing line), but no new pulsars were detected\footnote{One
of these sources is now known to harbor PSR~J1227$-$4853, which
recently transitioned to a radio-emitting state \citep[][and see
Table~\ref{tab:survey2}]{rrb+15}.}.

\begin{deluxetable*}{l@{\hspace{-1.5cm}}cccrrrlc}
\tabletypesize{\scriptsize}
\tablewidth{0.99\linewidth}
\tablecaption{\label{tab:survey2} Radio Searches of FGL Sources at Parkes: Source Information }
\tablecolumns{9}
\tablehead{
\colhead{3FGL name\tablenotemark{a}}       &
\colhead{$\Delta \theta$\tablenotemark{b}} &
\colhead{$r95$\tablenotemark{c}}           &
\colhead{Class\tablenotemark{d}}           &
\colhead{Sig\tablenotemark{e}}             &
\colhead{Curve\tablenotemark{f}}           &
\colhead{Var\tablenotemark{g}}             &
\colhead{Spectrum}                         &
\colhead{$N_{\rm obs}$\tablenotemark{i}}   \\
\colhead{}                              &
\colhead{(deg)}                         &
\colhead{(deg)}                         &
\colhead{}                              &
\colhead{($\sigma$)}                    &
\colhead{($\sigma$)}                    &
\colhead{}                              &
\colhead{notes\tablenotemark{h}}        &
\colhead{}
}
\startdata
{\bf J0101.0$-$6422}\tablenotemark{j} & 0.03 & 0.04 & PSR  & 26.8 &  6.8 &  55 & 1 CpR & 1 \\
\sout{J0133.0$-$4413} & 0.12 & 0.09 & bll  &  7.7 &  0.5 &  48 & 4 lhr  & 6 \\
\sout{J0216.1$-$7016} & 0.44 & 0.12 & bcu  &  6.0 &  0.5 &  49 & 4 ld   & 1 \\
\sout{J0602.8$-$4016} & 0.10 & 0.04 & bcu  & 15.3 &  1.8 &  49 & 4 lihr & 2 \\
\sout{J0744.8$-$4028} & 0.04 & 0.08 & bcu  &  7.4 &  1.9 &  39 & 3 ld   & 1 \\
\sout{J0802.3$-$5610} & 0.11 & 0.10 &      & 12.3 &  3.6 &  44 & 4 ld   & 5 \\
      J0933.9$-$6232  & 0.01 & 0.04 &      & 22.0 &  8.8 &  59 & 1 CPr  & 2 \\
      J0940.6$-$7609  & 0.01 & 0.10 &      & 10.0 &  1.1 &  48 & 3 ld   & 1 \\
\sout{J0940.7$-$6102} & 0.05 & 0.21 & bcu  &  6.4 &  2.7 &  51 & 3 pr   & 1 \\
      J0954.8$-$3948  & 0.04 & 0.09 &      & 18.8 &  3.9 &  51 & 3 D*   & 5 \\
{\bf J0955.6$-$6148}\tablenotemark{k} & 0.00 & 0.11 & psr  &  7.0 &  2.2 &  39 & 2 cr & 2 \\
{\bf J1012.0$-$4235}\tablenotemark{k} & 0.02 & 0.07 & psr  &  8.8 &  3.6 &  38 & 3 ?pr & 1 \\
      J1025.1$-$6507  & 0.02 & 0.09 &      &  7.3 &  2.7 &  52 & 3 ld   & 1 \\
{\bf J1035.7$-$6720}\tablenotemark{l} & 0.06 & 0.04 &      & 30.4 &  8.3 &  47 & 1 CpR & 3 \\
{\bf J1036.0$-$8317}\tablenotemark{k} & 0.01 & 0.12 & psr  &  6.6 &  2.3 &  42 & 3 ?p & 3 \\
\sout{J1057.7$-$6624} & 0.10 & 0.06 &      &  9.1 &  1.2 &  29 & 4 LD   & 4 \\
\sout{J1136.6$-$6826} & 0.01 & 0.11 & bcu  &  7.7 &  0.7 &  43 & 4 LD   & 1 \\
{\bf J1227.9$-$4854}\tablenotemark{m} & 0.04 & 0.04 & psr  & 38.6 &  4.6 &  74 & 3 Dh* & 2 \\
      J1231.6$-$5113  & 0.01 & 0.11 &      & 13.4 &  5.6 &  46 & 1 CPr  & 1 \\
\sout{J1238.3$-$4543} & 0.01 & 0.08 & bcu  &  8.0 &  0.3 &  61 & 4 ?lh  & 1 \\
\sout{J1306.8$-$4031} & 0.01 & 0.07 &      & 14.9 &  0.3 &  43 & 4 lD   & 1 \\
{\bf J1311.8$-$3430}\tablenotemark{n} & 0.02 & 0.02 & PSR  & 62.3 &  9.0 &  52 & 1 CR & 4 \\
      J1325.2$-$5411  & 0.00 & 0.10 &      &  7.0 &  2.5 &  31 & 3 ?ld  & 1 \\
      J1326.7$-$4727  & 0.02 & 0.05 & glc  & 11.3 &  6.1 &  66 & 1 Cpr  & 2 \\
      J1417.5$-$4402  & 0.00 & 0.06 &      & 12.5 &  1.4 &  54 & 2 cd   & 1 \\
\sout{J1417.7$-$5026} & 0.02 & 0.11 &      &  6.1 &  0.4 &  40 & 4 Ld   & 1 \\
{\bf J1514.2$-$4947}\tablenotemark{j} & 0.03 & 0.02 & PSR  & 39.4 &  9.3 &  35 & 1 CpR & 1 \\
      J1518.2$-$5232  & 0.04 & 0.07 &      & 10.3 &  3.6 &  62 & 2 cr   & 4 \\
{\bf J1536.3$-$4949}\tablenotemark{o} & 0.02 & 0.02 & psr  & 60.0 &  8.5 &  51 & 3 pRh & 6 \\
      J1539.2$-$3324  & 0.02 & 0.04 &      & 19.2 &  9.6 &  57 & 2 cPr  & 4 \\
\sout{J1603.7$-$6011} & 0.01 & 0.09 &      &  5.0 &  3.6 &  50 & 4 ?h   & 1 \\
\sout{J1617.4$-$5846} & 0.00 & 0.08 & fsrq & 15.8 &  1.6 & 128 & 5 LDV  & 1 \\
{\bf J1624.2$-$4041}\tablenotemark{l} & 0.02 & 0.04 &      & 19.2 &  7.3 &  50 & 1 cpR  & 2 \\
{\bf J1658.4$-$5323}\tablenotemark{j} & 0.10 & 0.06 & PSR  & 16.9 &  6.5 &  38 & 1 Cr & 1 \\
      J1702.8$-$5656  & 0.05 & 0.04 &      & 28.8 &  6.1 &  58 & 3 rdh* & 4 \\
\sout{J1717.4$-$5157} & 0.02 & 0.09 & fsrq & 12.2 &  2.5 & 383 & 5 LDV  & 1 \\
      J1736.2$-$4444  & 0.00 & 0.05 & glc  & 16.9 &  3.3 &  46 & 2 cd   & 1 \\
{\bf J1744.1$-$7619}\tablenotemark{l} & 0.01 & 0.03 &      & 32.8 &  9.9 &  51 & 1 CPR & 1 \\
{\bf J1747.6$-$4037}\tablenotemark{j} & 0.05 & 0.07 & PSR  & 10.0 &  2.8 &  46 & 2 ?c & 1 \\
      J1753.6$-$4447  & 0.01 & 0.08 &      & 11.1 &  4.0 &  41 & 1 cpr  & 1 \\
      J1803.3$-$6706  & 0.01 & 0.08 &      & 10.4 &  1.8 &  43 & 3 ldh  & 1 \\
      J1808.3$-$3357  & 0.03 & 0.09 &      &  8.7 &  4.4 &  50 & 1 cpr  & 4 \\
      J1831.6$-$6503  & 0.02 & 0.10 &      &  8.5 &  4.5 &  35 & 2 cPr  & 1 \\
{\bf J1902.0$-$5107}\tablenotemark{j} & 0.04 & 0.04 & PSR  & 28.9 &  6.2 &  50 & 1 CR & 1 \\
{\bf J1903.6$-$7052}\tablenotemark{k} & 0.08 & 0.05 & PSR  & 16.3 &  1.9 &  52 & 2 cd & 1 \\
{\bf J1946.4$-$5403}\tablenotemark{k} & 0.01 & 0.06 &      & 20.2 &  6.8 &  39 & 1 pr & 6 \\
\sout{J1959.8$-$4725} & 0.02 & 0.03 & bcu  & 19.2 &  3.9 &  49 & 4 pihr & 5 \\
      J2039.6$-$5618  & 0.04 & 0.04 &      & 25.3 &  5.1 &  34 & 1 CpR  & 4 \\
      J2043.8$-$4801  & 0.01 & 0.08 &      &  9.5 &  3.0 &  35 & 2 cr   & 2 \\
      J2112.5$-$3044  & 0.02 & 0.04 &      & 30.1 &  7.6 &  51 & 1 CpR  & 6 \\
      J2131.1$-$6625  & 0.01 & 0.11 &      & 10.1 &  3.1 &  52 & 2 rl   & 2 \\
      J2133.0$-$6433  & 0.05 & 0.10 &      & 10.1 &  4.7 &  49 & 2 Pr   & 7 \\
      J2200.0$-$6930  & 0.02 & 0.13 &      &  9.5 &  0.6 &  41 & 3 ld   & 1 \\
\sout{J2220.6$-$6833} & 0.02 & 0.11 &      &  5.7 &  1.1 &  50 & 4 lh   & 3 \\
{\bf J2241.6$-$5237}\tablenotemark{p} & 0.03 & 0.03 & PSR  & 51.7 & 12.6 &  60 & 1 CpR & 1 \\
      J2333.0$-$5525  & 0.00 & 0.08 &      & 10.7 &  2.5 &  42 & 2 cr   & 2
\enddata
\tablenotetext{a}{Boldfaced names denote 17 3FGL sources with now
known associated radio and/or gamma-ray pulsars; struck-through
names indicate 16 sources that we believe are no longer viable
pulsar candidates (see next-to-last column). }
\tablenotetext{b}{Offset between 3FGL position and pointing position
in Table~\ref{tab:survey} (where there are two of the latter, only
the second is listed here); the Parkes beam half-width at half-maximum
is 0.12\arcdeg. }
\tablenotetext{c}{Size of 3FGL source error box \citep[95\% confidence
level semi-major axis; all 3FGL properties listed here are taken
from the 3FGL catalog,][]{aaa+15}.  }
\tablenotetext{d}{Classification from 3FGL pipeline.  ``PSR'' is a
pulsar with LAT pulsations; ``psr'' is a positionally coincident
pulsar so far without LAT pulsations; ``bll'' is a BL Lac; ``bcu''
is an unclassified blazar; ``glc'' is a globular cluster; ``fsrq''
is a flat-spectrum radio quasar. }
\tablenotetext{e}{3FGL source significance. }
\tablenotetext{f}{Significance of curvature of 3FGL source spectrum
when fit to a log-parabolic model. }
\tablenotetext{g}{Variability index of source ($>73$ indicates
variability at the $>99$\% C.L.). }
\tablenotetext{h}{See Section~\ref{sec:unidclass} for a description
of these classifications and characteristics. }
\tablenotetext{i}{Number of observations of each target at position
closest to 3FGL source (from Table~\ref{tab:survey}), $\Delta \theta$
offset from it. }
\tablenotetext{j}{Radio MSP discoveries from our Parkes survey,
first reported in \citet{kcj+12}. }
\tablenotetext{k}{First reported in this work. }
\tablenotetext{l}{Pulsar discovered via gamma-ray pulsations
(H.~J.\ Pletsch, private communication). }
\tablenotetext{m}{Intermittently radio-emitting MSP \citep{rrb+15}. }
\tablenotetext{n}{MSP in 93-minute orbit discovered via gamma-ray
pulsations \citep{pgf+12}, subsequently detected at the GBT
\citep{rrc+13}. }
\tablenotetext{o}{Radio MSP discovered at the GMRT \citep{rap+12}. }
\tablenotetext{p}{Radio and gamma-ray MSP, discovered at Parkes
\citep{kjr+11}. }
\end{deluxetable*}

The 1FGL catalog was based on 11 months of LAT data.  The 2FGL
catalog \citep{naa+12} was based on two years of data.  In 2012 we
selected new search targets, based on a three-year source list
developed by the LAT collaboration but never published \citep[the
subsequent 3FGL catalog is based on four years of data and a different
pipeline;][]{aaa+15}.  As for the 1FGL searches \citep{kcj+12}, we
restricted ourselves to non-variable southern sources with no
plausible known blazar counterparts and with a LAT positional
uncertainty (95\% confidence level error radius) $\leq$7\arcmin,
to fit within the 1.4\,GHz primary beam of the Parkes telescope.
The remaining sources were classified by visual inspection of the
gamma-ray spectral energy distribution to pick out candidates with
spectral shape resembling those of known pulsars, which typically
have exponentially cut-off power-law spectra \citep{aaa+13}.  Spectral
modeling and source localization is more difficult for LAT sources
near the Galactic plane, and we only considered those with
$|b|>4\arcdeg$.

The new target set consisted of 49 sources, each observed between
one and seven times, for 122 integrations in the aggregate. Seven
of the 49 targets had also been observed in their prior 1FGL
incarnation (italicized in Table~\ref{tab:survey}). Among the
remaining 42, we discovered five MSPs (Table~\ref{tab:survey}).

Overall, we searched 56 individual LAT sources in our Parkes survey
using the analog filterbank system. We detected 11 MSPs, of which
10 were discoveries (Table~\ref{tab:survey}). In fact, radio and/or
gamma-ray pulsars are now known in 17 of those 56 sources
(Table~\ref{tab:survey2}). We discuss the sources, in particular
which might still be good pulsar candidates, in Section~\ref{sec:disc}.

Three of the five new MSPs were detected unbiasedly in all their
search observations (i.e., without prior knowledge of their DM or
approximate spin period). However, \psrj\ was detected in only three
of six observations, and \psrh\ was detected in only the third of
three search observations. We now consider the reasons behind these
failures to consistently detect some new pulsars.

\subsection{Sensitivity and Selection Effects} \label{sec:sensitivity}

\begin{figure}
\begin{center}
\includegraphics[scale=0.40]{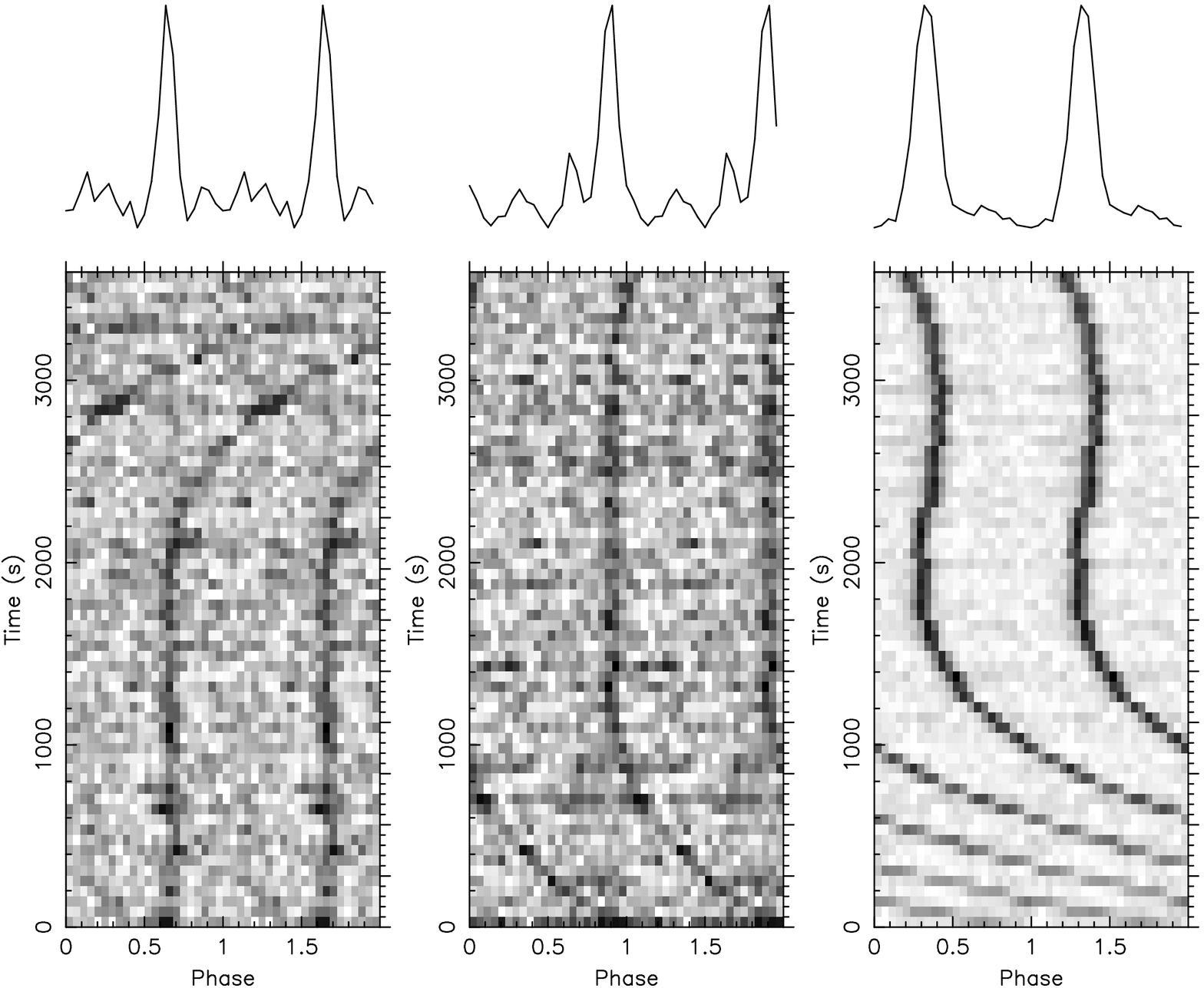}
\end{center}
\caption{\label{fig:J1946}
Pulse profiles (repeated twice) as a function of time during the
1\,hr observation times, and integrated (at top), for each of the
three unbiased detections of \psrj, folded according to the best
values of spin period and period derivative determined by the search
software. In the other three search observations of this source
(Table~\ref{tab:survey}) the pulsar was not detected unbiasedly,
because of the dilution of power in the Fourier domain caused by
the rapid non-linear change in projected orbital velocity for this
3\,hr binary system. A constant-acceleration search such as the
ones we performed cannot correct for this effect well enough.  }
\end{figure}

The nominal sensitivity of our survey to a $P=2$\,ms pulsar with a
duty cycle of 25\% and $\mbox{DM}\la40$\,pc\,cm$^{-3}$, for a typical
integration time of 1\,hr (Table~\ref{tab:survey}), and for a sky
temperature corresponding to the average at the locations of the
targets searched, was 0.2\,mJy --- provided that the dilution of
power in the Fourier domain caused by orbital motion in a putative
binary was ideally corrected by our acceleration search analysis.
However, two of the newly discovered MSPs have orbital periods $P_b$
such that the discovery integrations were 0.1--0.3\,$P_b$
(Section~\ref{sec:timing}), for which this idealized correction
breaks down badly; how badly, for a given system, depends on the
observed orbital phases \citep[see Figure~\ref{fig:J1946},
and][]{jk91,blw13}.

As well, a considerable fraction of LAT-selected MSPs is in eclipsing
binary systems \citep[e.g.,][]{hrm+11}. Multiple observations of a
promising source can thus help combat these factors affecting
detectability.  In addition, the flux density received from many
MSPs is severely affected by interstellar scintillation \citep[see,
e.g., Figure~1 of][]{lbb+13}. As we now show, this has played a
very significant role in our survey, and the early realization of
the magnitude of some of these selection effects is what led us to
do multiple observations of many unidentified LAT sources starting
in 2010 (Section~\ref{sec:2fgl}).

\subsubsection{Interstellar Scintillation and Detection Statistics}
\label{sec:flux}

\begin{figure*}[t]
\begin{center}
\includegraphics[scale=0.80]{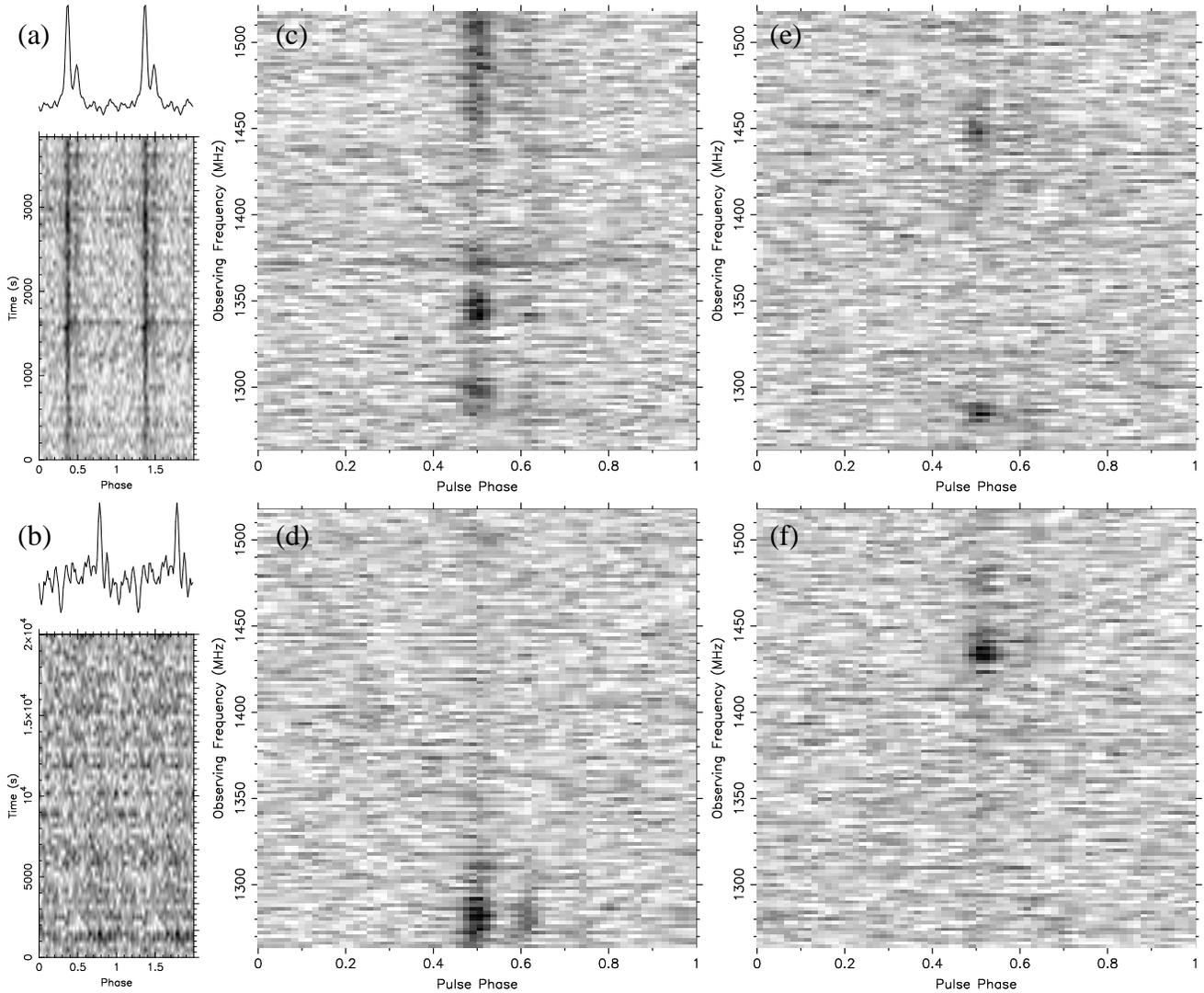}
\end{center}
\caption{\label{fig:scint}
\psra\ and its variable flux density due to scintillation in the
ISM, observed at Parkes at 1.4\,GHz with 256\,MHz of bandwidth using
the analog filterbank system.  All profiles are shown with 64 phase
bins per period ($P=3.6$\,ms), with a linear greyscale, and with
arbitrary phases between plots.  \textbf{(a)} The best detection
among 79 observations spanning 2 years, showing the signal strength
gently increasing during the observation, with the integrated profile
at the top. \textbf{(b)} An extremely weak detection, obtained only
by folding 6\,hr of data modulo the period predicted from the timing
solution.  \textbf{(c)} Signal strength as a function of radio
frequency for the observation also shown in (a).  \textbf{(d)} A
0.7\,hr good detection with average flux density half of that shown
in (a) and (c), and with very different time-integrated frequency
structure.  \textbf{(e, f)} Two observations separated by 9\,hr (of
length 1.0\,hr and 1.2\,hr, respectively), showing very different
frequency structure.  In many instances, one or two features of
$\approx 10$\,MHz in width dominate the detected signal.  Often,
no such feature is present and the pulsar is extremely faint, or
not detectable.  }
\end{figure*}

In order to acquire the detections required to determine the timing
solution for \psra\ (Section~\ref{sec:timing}), we used 100\,hr of
Parkes telescope time in 79 observations on 54 days spread over
2\,yr.  In only 27 of those observations did we detect the pulsar
in a {\em relatively\/} unbiased manner --- by dedispersing the raw
data using the known DM, performing an acceleration search, and
looking for a signal with period 3.589\,ms.  In several of those
observations the pulsar was not detectable without prior knowledge
of the DM and approximate period.  A pulsar like \psra\ is therefore
discoverable in a search like ours at Parkes less than 1/3 of the
time!  This is due to its small intrinsic flux density combined
with very large modulation owing to propagation through the dynamic
and inhomogeneous interstellar medium (see Figure~\ref{fig:scint}).

This large modulation of observed flux density is primarily caused
by strong diffractive interstellar scintillation. This causes the
detected pulsar signal strength in the time--frequency plane to
form patches (or ``scintles'') of characteristic size $\Delta \nu$
and $\Delta t$.  For a given observing frequency, $\Delta \nu$
increases strongly with decreasing pulsar distance $d$, while $\Delta
t$ increases particularly with decreasing pulsar velocity in the
plane of the sky $V_\perp$ \cite[see, e.g.,][]{jnk98,nnd+01,lk05}.
When $\Delta \nu$ or $\Delta t$ are large compared to the observing
bandwidth and integration time, deep fluctuations in observed flux
density result.  While we have not measured $\Delta \nu$ or $\Delta
t$ for \psra, the observations (e.g., Figure~\ref{fig:scint}) are
consistent with large values for both (e.g., $\ga 100$\,MHz and
$\ga 1$\,hr, respectively), accounting for the very large observed
modulations.

The measured flux density of \psrb\ is affected by scintillation
even more, ranging over a factor $>50$.  It was detected in a
relatively unbiased manner only 70\% of the time, and less than 1/2
of the time in the absence of prior DM and period information.
\psre, another MSP discovered in the first phase of our survey, for
which we have good statistics, is detectable 80\% of the time
\citep{kcj+12}. These detection statistics are based on typical
1\,hr individual Parkes timing observations (Section~\ref{sec:timing}).

For \psri, the observed flux densities range over a factor $>10$.
\psrd\ varies less: its recorded flux density has a standard deviation
of 40\% of the mean. These MSPs are reasonably bright, with relatively
narrow pulse profiles, and were always detected in typical 20\,min
Parkes timing observations.  For \psrc, with a large $\mbox{DM} =
153$\,pc\,cm$^{-3}$, the effects of scintillation modulate the flux
density that we record by only 25\% about the mean, and it was
detected 100\% of the time at Parkes and the GBT
(Section~\ref{sec:timing}).

Away from the Galactic plane, putative pulsar counterparts to
unidentified gamma-ray sources are largely expected to be relatively
nearby MSPs, which as a class have relatively small space velocities
\citep[e.g.,][]{gsf+11}.  Depending on observing parameters (frequency,
bandwidth, and integration time), these characteristics can make
them particularly susceptible to deep flux density fluctuations,
in turn with important implications for radio searches.  Scintillation
modulations in the strong regime have exponential statistics
\citep[e.g.,][]{ric90}, with median flux density measurements less
than the mean.  Two of the six MSPs for which we have already
obtained timing solutions (Section~\ref{sec:timing}) display flux
densities that vary by a factor of $>40$, with small median values
and, given scintillation statistics, not detectable in an unbiased
manner most of the time for the parameters of our Parkes survey.

The foregoing strongly suggests that an unidentified LAT source
that is a good pulsar candidate should be searched repeatedly before
much of a statement can be made about the likelihood of it being a
radio MSP beaming towards the Earth. Eight of the 10 MSPs that we
discovered were detected on their first observations
(Table~\ref{tab:survey}), but as the preceding account illustrates,
in some cases (particularly for PSRs~J1514$-$4946 and J1658$-$5324)
this was fortuitous.  Based on our empirical evidence, we judge
that every promising LAT source should be observed a {\em minimum\/}
of three or four times in a survey such as the one we did at Parkes
before it can reasonably be considered searched in an average sense.

As indicated in Table~\ref{tab:survey2}, we consider that a number
of our survey targets remain promising pulsar candidates, and at
least some should be re-searched in radio.  We discuss this more
in Sections~\ref{sec:unidclass} and \ref{sec:future}.

\subsection{Radio Timing} \label{sec:timing}

We began timing observations of all 10 MSPs immediately following
their discovery. In Table~\ref{tab:parms2} we present initial
parameters for the four most recent discoveries.  We have determined
phase-connected rotational ephemerides for the six remaining MSPs.
That for \psre\ was reported in \citet{kcj+12}; the other five are
given in Tables~\ref{tab:parms} and \ref{tab:J1903}, which also
list available flux density measurements.

\begin{deluxetable*}{lllll}
\tabletypesize{\footnotesize}
\tablewidth{0.99\linewidth}
\tablecaption{\label{tab:parms2} Preliminary Parameters for Four
Millisecond Pulsars }
\tablecolumns{5}
\tablehead{
{}              &
\colhead{\psrf} &
\colhead{\psrg} &
\colhead{\psrh} &
\colhead{\psrj}
}
\startdata
Right ascension\tablenotemark{a}, R.A. (J2000.0)\dotfill 
  & $09^{\rm h}55^{\rm m}39^{\rm s}$
  & $10^{\rm h}12^{\rm m}07^{\rm s}$
  & $10^{\rm h}36^{\rm m}20^{\rm s}$
  & $19^{\rm h}46^{\rm m}24^{\rm s}$
  \\
Declination\tablenotemark{a}, decl. (J2000.0)\dotfill
  & $-61\arcdeg 48\arcmin 36\arcsec$
  & $-42\arcdeg 35\arcmin 01\arcsec$
  & $-83\arcdeg 17\arcmin 01\arcsec$
  & $-54\arcdeg 02\arcmin 46\arcsec$
  \\
Spin period, $P$ (ms)\dotfill
  & 1.999
  & 3.101
  & 3.408
  & 2.710
  \\
Dispersion measure, DM (pc\,cm$^{-3}$)\dotfill
  &  160.7
  &   71.6
  &   27.0
  &   23.7
  \\
Orbital period, $P_b$ (d)\dotfill
  & 24.578
  & 37.972
  &  0.335
  &  0.130
  \\
Projected semi-major axis, $x$ (l-s)\dotfill
  & 13.283
  & 21.263
  &  0.506
  &  0.0435
  \\
Eccentricity, $e$\dotfill
  & 0.11
  & $<0.001$
  & $<0.001$ 
  & $<0.001$
  \\
Companion mass\tablenotemark{b}, $m_2$ (\msun)\dotfill
  & $>0.21$ 
  & $>0.26$
  & $>0.14$
  & $>0.021$
  \\
Galactic longitude, $l$ ($\deg$)\dotfill 
  & 283.7 
  & 274.2
  & 298.9
  & 343.9
  \\
Galactic latitude, $b$ ($\deg$)\dotfill  
  &  $-$5.7 
  &    11.2
  & $-$21.5
  & $-$29.6
  \\
DM-derived distance\tablenotemark{c}, $d$ (kpc)\dotfill
  & 3.8
  & 2.5
  & 1.0
  & 0.9
\enddata
\tablecomments{The listed $P$ and DM values are from the discovery
observations. Orbital parameters are from fits to sets of Doppler-shifted
barycentered periods. }
\tablenotetext{a}{These discovery pointing positions have $\pm7'$
uncertainties. On the assumption that the MSPs are associated with
the target gamma-ray sources, in some cases the position is better
constrained (see Table~\ref{tab:survey2}).  }
\tablenotetext{b}{Derived from the pulsar mass function $f_1$
assuming $m_1 = 1.35$\,\msun\ \citep{tc99} and $i<90\arcdeg$.  $f_1
= x^3 (2\pi/P_b)^2 T_\odot^{-1} = (m_2\sin i)^3/(m_1+m_2)^2$, where
$T_\odot \equiv G M_\odot/c^3 = 4.925\,\mu$s, $m_1$ and $m_2$ are
the pulsar and companion masses, respectively, and $i$ is the orbital
inclination angle.  }
\tablenotetext{c}{Using the \citet{cl02} Galactic free electron
density model. The individual estimates have substantial
uncertainties. }
\end{deluxetable*}

\begin{deluxetable*}{lllll}
\tabletypesize{\footnotesize}
\tablewidth{0.99\linewidth}
\tablecaption{\label{tab:parms} Parameters for Four Millisecond
Pulsars with Coherent Timing Solutions }
\tablecolumns{5}
\tablehead{
{}              &
\colhead{\psra} &
\colhead{\psrb} &
\colhead{\psrc} &
\colhead{\psrd}
}
\startdata
\cutinhead{Timing parameters}
Right ascension, R.A. (J2000.0)\dotfill 
  & $15^{\rm h}14^{\rm m}19\fs1141(1)$
  & $16^{\rm h}58^{\rm m}39\fs34359(9)$
  & $17^{\rm h}47^{\rm m}48\fs71692(3)$
  & $19^{\rm h}02^{\rm m}02\fs84821(9)$
  \\
Declination, decl. (J2000.0)\dotfill
  & $-49\arcdeg 46\arcmin 15\farcs516(5)$
  & $-53\arcdeg 24\arcmin 07\farcs003(1)$
  & $-40\arcdeg 36\arcmin 54\farcs773(1)$
  & $-51\arcdeg 05\arcmin 56\farcs9695(8)$
  \\
Proper motion in R.A., $\dot \alpha \cos (\delta)$ (mas\,yr$^{-1}$)\dotfill
  & $-$0.3(32)
  &    0.2(8)
  & $-$0.8(6)
  & $-$4.8(13)
  \\
Proper motion in decl., $\dot \delta$ (mas\,yr$^{-1}$)\dotfill
  & $-$30.0(66)
  &     4.9(23)
  &  $-$4.9(16)
  &  $-$4.4(16)
  \\
Spin frequency, $f$ (Hz)\dotfill
  & 278.60300920296(5)
  & 409.95436264371(4)
  & 607.67753906573(2)
  & 573.92104496683(5)
  \\
Frequency derivative, $\dot f$ (Hz\,s$^{-1}$)\dotfill 
  & $-1.4473(8)\times10^{-15}$
  & $-1.8746(6)\times10^{-15}$
  & $-4.8510(5)\times10^{-15}$
  & $-3.0301(4)\times10^{-15}$ 
  \\
Epoch (MJD)\dotfill
  & 55520.0
  & 55520.0
  & 55520.0
  & 55520.0
  \\
Dispersion measure, DM (pc\,cm$^{-3}$)\dotfill
  &  31.05(2)
  &  30.81(3)  
  & 152.98(1)
  &  36.25(1)
  \\
Orbital period, $P_b$ (d)\dotfill
  & 1.922653523(5)
  & \nodata
  & \nodata
  & 2.0118037388(9)
  \\
Projected semi-major axis, $x$ (l-s)\dotfill
  & 1.933268(2)
  & \nodata
  & \nodata
  & 1.9019570(7)
  \\
Time of ascending node, $T_{\rm asc}$ (MJD)\dotfill 
  & 55585.8605555(3)
  & \nodata
  & \nodata
  & 55162.2815604(1)
  \\
$e \sin \omega$\tablenotemark{a}, EPS1\dotfill
  & $6.453587(2)\times10^{-6}$
  & \nodata
  & \nodata
  & $5.5239429(7)\times10^{-6}$
  \\
$e \cos \omega$\tablenotemark{a}, EPS2\dotfill
  & $8.789469(3)\times10^{-6}$
  & \nodata
  & \nodata
  & $-1.9671264(9)\times10^{-6}$
  \\
Span of timing data (MJD)\dotfill
  & 55160--57013 
  & 55166--57057 
  & 55161--56993 
  & 55161--57014 
  \\
rms timing residual ($\mu$s)\dotfill  
  & 8.3
  & 2.9
  & 2.4
  & 3.8
  \\
[-2pt]
\cutinhead{Flux densities\tablenotemark{b} and rotation measures}
0.8\,GHz flux density, $S_{0.8}$ (mJy)\dotfill
  & \nodata
  & \nodata
  & 4.8
  & \nodata
  \\
1.4\,GHz flux density, $S_{1.4}$ (mJy)\dotfill
  & $0.17\pm0.11$ ($N=71$)
  & $0.7\pm0.5$ ($N=37$)
  & \nodata
  & $1.2\pm0.5$ ($N=69$)
  \\
1.5\,GHz flux density, $S_{1.5}$ (mJy)\dotfill
  & \nodata
  & \nodata
  & 0.9
  & \nodata
  \\
2\,GHz flux density, $S_{2}$ (mJy)\dotfill
  & \nodata
  & \nodata
  & $0.5\pm0.1$ ($N=28$)
  & \nodata
  \\
Rotation measure, RM (rad\,m$^{-2}$)\dotfill
  & $35\pm15$
  & $4\pm7$
  & $-39\pm2$
  & \nodata
  \\
[-2pt]
\cutinhead{Derived parameters\tablenotemark{c}}
Spin period, $P$ (ms)\dotfill
  & 3.589
  & 2.439
  & 1.645
  & 1.742
  \\
Characteristic age, $\tau_c$ ($10^9$\,yr)\dotfill
  & 4.9
  & 3.5
  & 2.0
  & 3.1
  \\
Spin-down luminosity, $\dot E$ ($10^{34}$\,erg\,s$^{-1}$)\dotfill
  & 1.0
  & 3.0
  & 11.3
  & 6.7
  \\
Surface dipole magnetic field strength, $B$ ($10^8$\,G) \dotfill
  & 2.1
  & 1.7
  & 1.5
  & 1.3
  \\
Eccentricity\tablenotemark{a}, $e$\dotfill
  & $(1.1\pm0.2)\times 10^{-5}$
  & \nodata
  & \nodata
  & $(5.9\pm0.7)\times 10^{-6}$
  \\
Mass function, $f_1$ (\msun)\dotfill
  & 0.00210
  & \nodata
  & \nodata
  & 0.00183
  \\
Companion mass, $m_2$ (\msun)\dotfill
  & $>0.17$ 
  & \nodata
  & \nodata
  & $>0.16$ 
  \\
Spectral index\tablenotemark{d}, $\alpha$\dotfill
  & \nodata
  & \nodata
  & $\approx -2.5$
  & \nodata
  \\
Galactic longitude, $l$ ($\deg$)\dotfill 
  & 325.25
  & 334.87
  & 350.21
  & 345.65
  \\
Galactic latitude, $b$ ($\deg$)\dotfill  
  &     6.81
  &  $-$6.63
  &  $-$6.41
  & $-$22.38
  \\
DM-derived distance, $d$ (kpc)\dotfill
  & 0.9
  & 0.9
  & 3.4
  & 1.2
  \\
Composite proper motion, $\mu$ (mas\,yr$^{-1}$)\dotfill
  & $30.0\pm6.6$
  & $ 4.9\pm2.2$
  & $ 5.0\pm1.6$
  & $ 6.5\pm1.4$
  \\
Transverse velocity, $V_\perp$ (km\,s$^{-1}$)\dotfill
  & $\approx 130$
  & $\approx 20$
  & $\approx 80$
  & $\approx 40$
\enddata
\tablecomments{All uncertainties are reported at the $1\,\sigma$
level. Numbers in parentheses represent the TEMPO2 timing uncertainties
on the last digits quoted.  }
\tablenotetext{a}{The orbital eccentricities were derived from the
parameters $(e \sin \omega, e \cos \omega)$, fitted using the TEMPO2
ELL1 binary model \citep{lcw+01}.  }
\tablenotetext{b}{$S_{1.4}$ and $S_2$ values are averages and
standard deviations for $N$ detections.  Individual flux densities
were determined by computing the area under each pulse profile
compared to its off-pulse rms, scaled using the measured system
equivalent flux density at the location of the pulsar. Values for
\psrc\ are from GBT observations; $S_{0.8}$ and $S_{1.5}$ are from
single flux-calibrated observations (Figure~\ref{fig:pol}c) and
have uncertainties $\la 10\%$. For one similar observation at 2\,GHz,
$S_2=0.45$\,mJy.}
\tablenotetext{c}{The following have been used: $\tau_c = P/(2\dot
P)$, $\dot E = 4 \pi^2 \times 10^{45} \dot P/P^3$\,erg\,s$^{-1}$,
and $B = 3.2\times10^{19} (P\dot P)^{1/2}$\,G, with $P$ in s, where
$P=1/f$. The listed values of these parameters include corrections
for acceleration effects \citep[mainly due to proper motion;
see][]{ctk94}.  }
\tablenotetext{d}{$S_\nu \propto \nu^{\alpha}$, where $S_\nu$
is the flux density at frequency $\nu$.}
\end{deluxetable*}

\begin{deluxetable}{ll}
\tabletypesize{\footnotesize}
\tablewidth{0pt}
\tablecaption{\label{tab:J1903} Radio and Gamma-ray Parameters of \psri\ }
\tablecolumns{2}
\tablehead{
\colhead{Parameter} &
\colhead{Value}
}
\startdata
\cutinhead{Timing parameters}
Right ascension, R.A. (J2000.0)\dotfill 
  & $19^{\rm h}03^{\rm m}38\fs7935(3)$
  \\
Declination, decl. (J2000.0)\dotfill
  & $-70\arcdeg 51\arcmin 43\farcs461(2)$
  \\
Proper motion in R.A., $\dot \alpha \cos (\delta)$ (mas\,yr$^{-1}$)\dotfill
  & $-8.8(16)$
  \\
Proper motion in decl., $\dot \delta$ (mas\,yr$^{-1}$)\dotfill
  & $-16(2)$
  \\
Spin frequency, $f$ (Hz)\dotfill
  & 277.94006243351(8)
  \\
Frequency derivative, $\dot f$ (Hz\,s$^{-1}$)\dotfill 
  & $-8.06(4)\times10^{-16}$
  \\
Epoch (MJD)\dotfill
  & 56526.0
  \\
Dispersion measure, DM (pc\,cm$^{-3}$)\dotfill
  &  19.66(1)
  \\
Orbital period, $P_b$ (d)\dotfill
  & 11.05079833(2)
  \\
Projected semi-major axis, $x$ (l-s)\dotfill
  & 9.938869(2)
  \\
Time of ascending node, $T_{\rm asc}$ (MJD)\dotfill 
  & 56027.2292914(7)
  \\
$e \sin \omega$, EPS1\dotfill
  & $2.0261968(5)\times10^{-6}$
  \\
$e \cos \omega$, EPS2\dotfill
  & $1.1855635(5)\times10^{-7}$
  \\
Span of timing data\tablenotemark{a} (MJD)\dotfill
  & 54760--57013 
  \\
rms timing residual\tablenotemark{a} ($\mu$s)\dotfill  
  & 12.8
  \\
[-2pt]
\cutinhead{Flux densities\tablenotemark{b} and rotation measure}
1.4\,GHz flux density, $S_{1.4}$ (mJy)\dotfill
  & $\approx 0.6$ ($N=9$)
  \\
3.1\,GHz flux density, $S_{3.1}$ (mJy)\dotfill
  & $\sim 0.14$ ($N=3$)
  \\
Rotation measure, RM (rad\,m$^{-2}$)\dotfill
  & $11\pm24$
  \\ 
[-2pt]
\cutinhead{Derived parameters\tablenotemark{c}}
Spin period, $P$ (ms)\dotfill
  & 3.597
  \\
Characteristic age, $\tau_c$ ($10^9$\,yr)\dotfill
  & 7.1
  \\
Spin-down luminosity, $\dot E$ ($10^{33}$\,erg\,s$^{-1}$)\dotfill
  & 6.8
  \\
Surface dipole magnetic field strength, $B$ ($10^8$\,G) \dotfill
  & 1.7
  \\
Eccentricity, $e$\dotfill
  & $(2.0\pm0.5)\times 10^{-6}$
  \\
Mass function, $f_1$ (\msun)\dotfill
  & 0.00863
  \\
Companion mass, $m_2$ (\msun)\dotfill
  & $>0.28$ 
  \\
Spectral index, $\alpha$\dotfill
  & $\sim -1.8$
  \\
Galactic longitude, $l$ ($\deg$)\dotfill 
  & 324.39
  \\
Galactic latitude, $b$ ($\deg$)\dotfill 
  & $-$26.51
  \\
DM-derived distance, $d$ (kpc)\dotfill
  & 0.8
  \\
Composite proper motion, $\mu$ (mas\,yr$^{-1}$)\dotfill
  & $18.3\pm2.0$
  \\
Transverse velocity, $V_\perp$ (km\,s$^{-1}$)\dotfill
  & $\approx 70$
  \\
[-2pt]
\cutinhead{Gamma-ray parameters}
Gamma-ray--radio profile lag\tablenotemark{d}, $\delta$ ($P$)\dotfill 
  & $0.55\pm0.05$ 
  \\
Gamma-ray ($>0.1$\,GeV) photon index, $\Gamma$\dotfill
  & $1.90\pm0.16$
  \\
Gamma-ray cut-off energy, $E_c$ (GeV)\dotfill
  & $7.8\pm4.0$
  \\
Photon flux ($>0.1$\,GeV) ($10^{-8}\,{\rm cm^{-2}\,s^{-1}}$)\dotfill
  & $1.2\pm0.3$
  \\
Energy flux ($>0.1$\,GeV) ($10^{-11}\,{\rm erg\,cm^{-2}\,s^{-1}}$)\dotfill
  & $0.9\pm0.2$
\enddata
\tablecomments{All uncertainties are reported at the $1\,\sigma$
level. Numbers in parentheses represent the TEMPO2 timing uncertainties
on the last digits quoted. Uncertainties for gamma-ray spectral
parameters are statistical only \citep[for a discussion of systematic
errors, see][]{aaa+15}. }
\tablenotetext{a}{We have used both radio and gamma-ray TOAs to
derive this timing solution. }
\tablenotetext{b}{The listed values are median flux densities for
$N$ calibrated detections. Individual $S_{1.4}$ values ranged over
0.13--1.5\,mJy. In four 3.1\,GHz observations we did not detect the
pulsar.}
\tablenotetext{c}{The listed values of $\tau_c$, $\dot E$, and $B$
include corrections for acceleration effects. }
\tablenotetext{d}{This is measured from the profiles in
Figure~\ref{fig:J1903}, with the gamma-ray profile centroid at phase
$\phi=0.27$ and the radio profile reference phase mid-way between
its full observed span ($0.55 < \phi < 0.9$). }
\end{deluxetable}

\begin{figure*}[t]
\begin{center}
\includegraphics[scale=0.95]{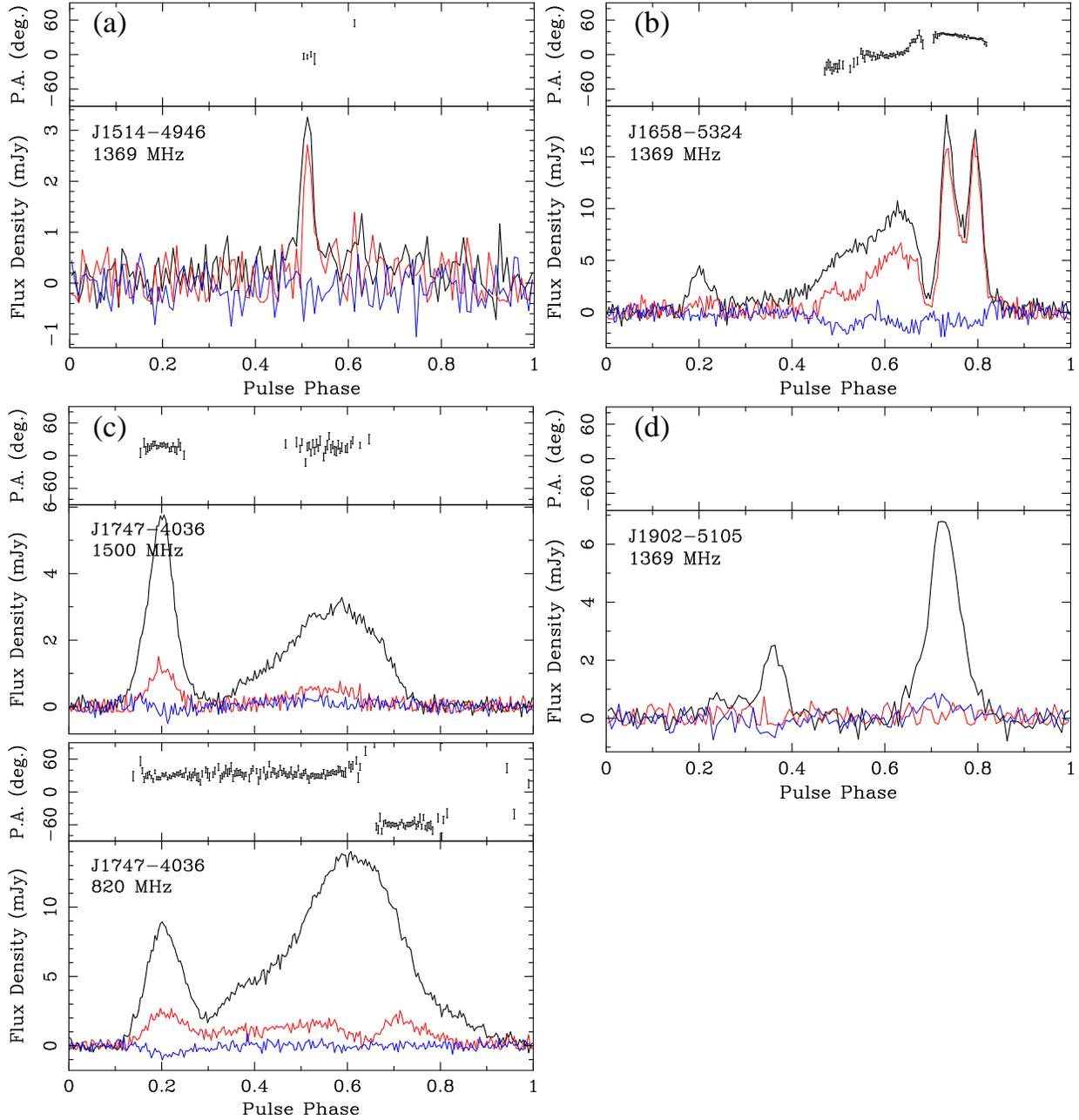}
\end{center}
\caption{\label{fig:pol}
Polarimetric pulse profiles for PSRs~J1514$-$4946 {\bf (a)},
J1658$-$5324 {\bf (b)}, J1747$-$4036 {\bf (c)}, and J1902$-$5105
{\bf (d)}. In the bottom sub-plots, the black line corresponds to
total intensity, while the red and blue traces represent linear and
circular polarization, respectively.  In the top sub-plots, the
position angle of linear polarization (P.A.) is plotted for bins
in which the linear signal-to-noise ratio $>3$, corrected to the
reference frame of the pulsar using the RMs listed in
Table~\ref{tab:parms}. \psrc\ was observed at the GBT with GUPPI
(recording a bandwidth of 800\,MHz at a central frequency of 1500\,MHz
and 200\,MHz of bandwidth at 820\,MHz, with the two profiles aligned
by eye). All others were observed at Parkes with PDFB3 recording
256\,MHz of bandwidth.}
\end{figure*}

\psrc\ was observed mainly at the NRAO Green Bank Telescope (GBT),
using the GUPPI
spectrometer\footnote{https://safe.nrao.edu/wiki/bin/view/CICADA/GUPPiUsersGuide.}
to sample a bandwidth of 800\,MHz centered at 2\,GHz, with typical
integration times of 5\,min.

The remaining MSPs were observed exclusively at Parkes where we
first used the analog filterbank/PMDAQ data acquisition system as
employed in the discovery observations, and later a digital filterbank
(PDFB3/4), in all cases centered at 1.4\,GHz.  Each observation
typically lasted for 1\,hr, except for the brighter PSRs~J1902$-$5105
and J1903$-$7051 (about 20\,min).  Apart from a dense set of
observations to obtain orbital parameters for the binary pulsars,
and to unambiguously establish pulse numbering, we aimed to detect
each pulsar approximately monthly over a span of 1--2\,yr.  Especially
for \psra, this required a very large number of observations, because
the pulsar is very faint on average and its received radio flux
density varies enormously due to interstellar scintillation
(Section~\ref{sec:flux}). We improved the precision of the timing
solutions by substantially extending the measurement baselines with
a few additional observations in 2014 and early 2015.

We used the PRESTO and PSRCHIVE \citep{hvm04} software to analyze
the raw data and obtain pulse times-of-arrival (TOAs).  For the
MSPs with sparse radio detections, we obtained initial estimates
of the orbital parameters using the method described in \citet{fkl01}.
We then used TEMPO\footnote{http://tempo.sourceforge.net.} and
TEMPO2 \citep{hem06} to obtain phase-connected timing solutions.
Starting with solutions spanning $\approx 1$\,yr for each pulsar,
we detected gamma-ray pulsations for six MSPs.  For \psri\ we then
obtained a few LAT TOAs.  While the radio TOAs have much higher
precision, they span only 3\,yr.  Adding gamma-ray TOAs increases
the solution span and improves some results, in particular the
proper motion measurement (see Section~\ref{sec:gtiming}).

\subsection{Polarimetry} \label{sec:pol}

In principle, study of the polarized radio emission from pulsars
can constrain the magnetic field geometry and, particularly when
considered together with gamma-ray profile characteristics, can
elucidate emission locations and mechanisms (cf.\
Section~\ref{sec:sixgamma}).

In order to measure the polarization characteristics of each pulsar,
we used the digital filterbank PDFB3 at Parkes, and GUPPI at GBT
in coherent dedispersion mode, to do a few observations recording
calibrated folded full-Stokes pulse profiles.  These data were
analyzed in standard fashion with PSRCHIVE.  The resulting profiles
are shown in Figures~\ref{fig:pol} and \ref{fig:J1903}, and rotation
measures (RMs) are listed in Tables~\ref{tab:parms} and \ref{tab:J1903}.

The main profile component of \psra\ is about 80\% linearly polarized,
with a flat position angle (P.A.) of linear polarization
(Figure~\ref{fig:pol}a).

\psrb\ has a complex profile, with a high (up to 100\%) linear
polarization fraction in most pulse components (Figure~\ref{fig:pol}b).
Such a high level of linear polarization is relatively uncommon
among MSPs \citep[see][]{ymv+11,dhm+15}.  Despite the P.A.\ being
measured over a large span of rotation phase, we were not able to
obtain a satisfactory rotating vector model fit \citep{rc69a}.

The profile of \psrc\ is curious.  This MSP has the second largest
value of $\mbox{DM}/P$ among those known in the Galactic disk, and
so was particularly well-suited for observing with a coherent
dedispersion system.  We show two such observations in
Figure~\ref{fig:pol}c.  At 1.5\,GHz, the profile has two featureless
components, with a very flat P.A.\ throughout.  Lack of variation
in P.A.\ across the entire profile is unusual \cite[there is no
such example in a well-studied sample of 24 MSPs;][]{dhm+15}.  The
profile observed at 2\,GHz (which overlaps in frequency with that
at 1.5\,GHz) looks similar.  At 0.8\,GHz, however (bottom panels
of Figure~\ref{fig:pol}c), the second component is brighter than
the first, signifying that it has a steeper spectrum, and a third
component appears (more easily discernible in linear polarization),
with P.A.s offset by about $90\arcdeg$ from the rest of the profile.
At this frequency, emission is detectable across 85\% of the pulse
period (but the observed profile may be somewhat affected by
multi-path propagation; Cordes \& Lazio 2002 predict pulse broadening
of $0.02\,P$)\nocite{cl02}.

The \psrd\ profile is relatively unusual in showing no discernible
linear polarization (there is possibly a very small amount of circular
polarization in both principal components; Figure~\ref{fig:pol}d).

The radio profile of \psri\ is shown in the bottom panel of
Figure~\ref{fig:J1903}. It displays a weak, highly-polarized component
(pulse phase $\phi\approx0.58$) leading a much stronger double-peaked
component ($0.65 \la \phi \la 0.85$).  The first peak is highly
linearly polarized and also shows a degree of circular polarization,
while the trailing (brightest) peak is completely unpolarized.

\subsection{Gamma-ray Observations and Analysis} \label{sec:gamma}

The gamma-ray properties of the first five MSPs we discovered (\psre\
and those in Table~\ref{tab:parms}) have been reported elsewhere
\citep{kcj+12,aaa+13}, and we focus here on the newly-detected
gamma-ray pulsations of \psri.  To characterize its properties, we
analyzed ``reprocessed Pass 7'' {\tt SOURCE} class \fermi{}-LAT
events collected between 2008 August 4 (the start of nominal
operations) and 2015 February 1, and with energies between 0.1 and
30\,GeV.  The data were filtered to exclude events whose reconstructed
direction exceeds a zenith angle of 100\arcdeg\ and those taken
when the observatory was rocked more than 52\arcdeg\ from the zenith,
or when the region of interest (ROI) around the pulsar approached
the Earth's limb.  These cuts minimize the bright gamma-ray background
contribution from the limb of the Earth.

\subsubsection{Spectral Analysis and Photon Weights} \label{sec:ga}

We considered a 20\arcdeg$\times$20\arcdeg\ region around \psri,
and used the binned likelihood formulation of {\em gtlike} (\fermi{}
Science Tools v.\
09-35-02\footnote{http://fermi.gsfc.nasa.gov/ssc/data/analysis/software.})
to fit an exponentially cut-off power law to the pulsar emission,
$dN/dE=N_0\,(E/E_0)^{-\Gamma}\,\exp(-E/E_c)$.  We used the
\texttt{P7REP\_SOURCE\_V15} model of the instrument response
functions\footnote{http://fermi.gsfc.nasa.gov/ssc/data/access/lat/BackgroundModels.html.}
and the isotropic and diffuse background models of the 3FGL catalog
\citep{aaa+15}, as well as the point sources therein.  The best-fit
parameters and their error estimates are listed in the last section
of Table~\ref{tab:J1903}.  With models for the pulsar and background
sources, we can compute a weight $w_i$ for each photon $i$ giving
the probability that the photon originated from the pulsar
\citep{ker11}.  These weights allow improved separation of the
pulsar signal from its background in the following analyses.

\subsubsection{Pulsar Timing} \label{sec:gtiming}

Folding LAT data for \psri\ based on the radio-only timing solution
results in a clear drift in the position of the gamma-ray profile
peak outside the time interval covered by the radio ephemeris.
Thus, using the same selection criterion as for the pulsar light
curve (Section~\ref{sec:glc}), we have extracted 16 gamma-ray TOAs
following the method of \citet{rkp+11} and used these to extend the
timing solution.  The resulting parameters, including an improved
measurement of the proper motion, appear in Table~\ref{tab:J1903}.

\begin{figure}
\begin{center}
\includegraphics[scale=0.47]{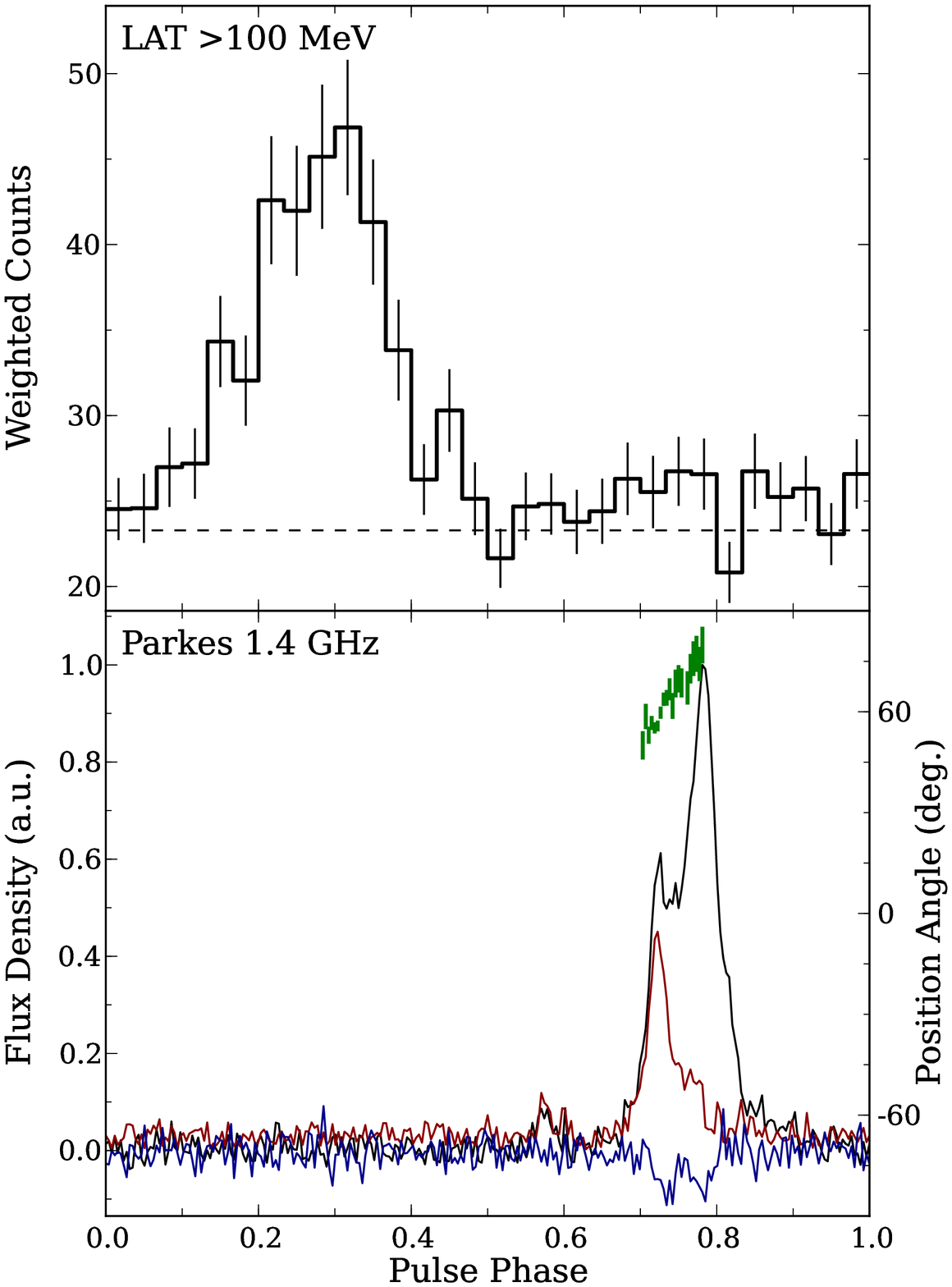}
\end{center}
\caption{\label{fig:J1903}
Gamma-ray and polarimetric radio profiles of \psri.  {\em Top:}
Weighted $>0.1$\,GeV light curve (6.5\,yr of LAT data), displayed
with 30 phase bins, including estimates of the error bars and the
background level (see Section~\ref{sec:glc}).  {\em Bottom:}
Phase-aligned Parkes 1.4\,GHz pulse profile (black: total intensity;
red: linear polarization; blue: circular polarization).  The green
error bars are centered on the position angles of linear polarization
(displayed for $\mbox{RM}=0$ with an offset of +130\arcdeg\ for
ease of view).  }
\end{figure}

\subsubsection{Light Curve} \label{sec:glc}

The gamma-ray light curve in the top panel of Figure~\ref{fig:J1903},
with H-test \citep{drs89,ker11a} value of 231, is a probability-weighted
histogram of the photon phases.  Because the probability weights
$w_i$ allow an optimal ``soft'' cut, no tuning of the photon selection
is required, and the histogram includes all events within 2\arcdeg\
of the pulsar position.  The structure of the light curve is robust
against additional background, so here, unlike in the spectral
analysis, we do not apply the cut on zenith angle when the ROI is
near the horizon, increasing the total number of photons by $\approx
25$\% and slightly increasing the background level.  Error bars for
a bin follow the typical prescription $\sigma_j^2 =
\sum_{i=1}^{N_{\gamma,j}} w_i^2$, with the sum over the $N_{\gamma,j}$
photons in the $j$th bin.  The background level is given by
$(\sum_{i=1}^{N_{\gamma}} w_i - \sum_i w_i^2) / N_{\rm bin}$, with
the sums over all $N_{\gamma}$ photons in the profile and $N_{\rm
bin}$ the total number of bins.  This level, based on the spectral
model, represents the expected contribution from all diffuse and
background point sources.  Both pulsed and unpulsed emission from
the position of \psri\ will show up as a signal in excess of this
background level.  However, from inspection of the light curve, it
is evident that \psri\ has no substantial unpulsed component.  A
slight excess may be due to an increased background over the spectral
model due to the less stringent zenith cut.

\subsection{X-ray Observations} \label{sec:xray}

\begin{deluxetable*}{llllll}
\tabletypesize{\footnotesize}
\tablewidth{0.99\linewidth}
\tablecaption{\label{tab:xray} X-ray Observations of Five
Millisecond Pulsars }
\tablecolumns{5}
\tablehead{
{}              &
\colhead{\psra} &
\colhead{\psrb} &
\colhead{\psrc} &
\colhead{\psrd} &
\colhead{\psri}
}
\startdata
Telescope/Instrument\dotfill 
  & \cxo/ACIS-S 
  & \cxo/ACIS-S
  & \swift/XRT~PC
  & \swift/XRT~PC
  & \swift/XRT~PC
  \\
Exposure (ks)\dotfill
  & 9.9 
  & 9.9
  & 3.3
  & 4.2
  & 3.2
  \\
Background-subtracted counts\dotfill
  &    9
  &    23
  &  $<6.6$\tablenotemark{a}
  & $<12.7$\tablenotemark{a}
  &  $<6.6$\tablenotemark{a}
  \\
$N_{\rm H}$\tablenotemark{b} ($10^{21}$\,cm$^{-2}$)\dotfill
  & 1
  & 1
  & 5
  & 1
  & 0.6
  \\
X-ray flux\tablenotemark{c}, $f_{\rm 0.1-2.4\,keV}$ ($10^{-14}$\,erg\,cm$^{-2}$\,s$^{-1}$)\dotfill
  &    1.1
  &    2.9
  & $<23$
  & $<12$\tablenotemark{d}
  &  $<7$
  \\
X-ray luminosity\tablenotemark{e}, $L_{\rm 0.1-2.4\,keV}$ ($10^{-4} \dot E$)\dotfill
  &    1.1
  &    0.9
  & $<28$
  &  $<3$
  &  $<8$
\enddata
\tablecomments{A blackbody model with $kT = 0.2$\,keV is assumed
in all cases, absorbed by the indicated column density $N_{\rm H}$.
Considering instead a power-law spectrum with photon index $\Gamma
= 2$, and calculating flux/luminosities for the 2--10\,keV range,
does not fundamentally alter our conclusions (Section~\ref{sec:xray}).  }
\tablenotetext{a}{We used $47\arcsec$-radius extraction regions
around each pulsar (90\% PSF radius for XRT), and $141\arcsec$ radii
to estimate backgrounds. For two sources we obtained zero
background-subtracted counts, while for \psrd\ we obtained three
counts.  In each instance we convert to a $3\,\sigma$ upper limit
\citep[see][]{geh86}. }
\tablenotetext{b}{Absorbing columns are estimated from the DMs
according to $N_{\rm H} (10^{21}\mbox{cm}^{-2}) = 0.03\,\mbox{DM}$
\citep[see][]{hnk13}. }
\tablenotetext{c}{We used http://heasarc.gsfc.nasa.gov/Tools/w3pimms.html
to obtain these unabsorbed flux estimates (or $3\,\sigma$ limits).  }
\tablenotetext{d}{\citet{tkn+12} report a somewhat lower flux limit
based on a 38\,ks {\em Suzaku} XIS observation. }
\tablenotetext{e}{Isotropic 0.1--2.4\,keV luminosities are given in
terms of the acceleration-corrected values of $\dot E$, and for the
DM-derived distances, listed in Tables~\ref{tab:parms} and
\ref{tab:J1903}. }
\end{deluxetable*}

In an attempt to assist in determining accurate positions for the
pulsars before timing solutions existed, we undertook X-ray
observations of the fields of PSRs~J1514$-$4946 and J1658$-$5324
with the ACIS-S camera on board the {\em Chandra X-ray Observatory}
(\cxo).  For both observations the (then) best pulsar positions
were centered on the back-illuminated S3 chip, and we analyzed the
event data with the standard CIAO version 4.7 software \citep{fma+06}.
After excluding events outside the energy range 0.3--7.0\,keV we
searched the fields for any point-like X-ray source possibly
associated with the pulsars using the {\tt celldetect} source search
tool.  In each case we detected a faint source $0\farcs3\pm0\farcs6$
away from the pulsar timing position. We name these, respectively,
CXOU~J151419$-$494615 and CXOU~J165839$-$532406, which we identify
as the pulsar counterparts on the basis of positional coincidence.
Observation and source parameters are listed in Table~\ref{tab:xray}.

For three other MSPs, \swift\ X-Ray Telescope (XRT) observations
have been undertaken as part of a \swift\ campaign of observations
of \fermi-LAT unassociated sources.  No sources were detected at
the pulsar positions. More information, including flux and luminosity
upper limits, is given in Table~\ref{tab:xray}.

All these X-ray detections and upper limits are consistent with the
known distribution of MSP X-ray luminosities \citep[e.g.,][and
references therein]{pccm02}.

\section{Discussion} \label{sec:disc} 

In our directed radio survey with the Parkes telescope of 56
unidentified \fermi-LAT gamma-ray sources, we detected 11 MSPs, 10
of them discoveries (Section~\ref{sec:searches}).

\subsection{Ten New Millisecond Pulsars}

Two of the 10 new pulsars are isolated (PSRs~J1658$-$5324 and
J1747$-$4036; Table~\ref{tab:parms}). Another five are in circular
orbits ($e<10^{-3}$) with 0.2--0.3\,\msun\ companions, likely
helium-core white dwarfs \citep[PSRs~J0101$-$6422, J1012$-$4235,
J1514$-$4946, J1902$-$5105, and J1903$-$7051;][and Tables~\ref{tab:parms2},
\ref{tab:parms} and \ref{tab:J1903}]{kcj+12}. Three of these have
orbital periods of about 2\,d, while the other two have periods of
11\,d and 38\,d.  These characteristics are typical of MSPs known
in the Galactic disk.  The timing precision of \psrc\ is good enough
that it has been added to the NANOGrav gravitational wave pulsar
timing array \citep{mac13}.

The remaining three systems (see Table~\ref{tab:parms2}) are less
common. \psrf\ is in a $P_b = 24$\,d orbit with a $m_2 \approx
0.25$\,\msun\ companion, but with a very significant eccentricity
($e = 0.11$). It joins four Galactic MSP systems with broadly similar
parameters: $0.2 < m_2 <0.3$\,\msun, $0.03 < e < 0.13$, $22 < P_b
< 32$\,d \citep[see][and references therein]{ant14,kls+15}. Such
eccentric systems are not predicted through the standard MSP formation
channels \citep[see][]{pk94}, leading to alternative scenarios
\citep{ft14,ant14}.

\psrh\ is in an 8\,hr orbit with a $\approx 0.16$\,\msun\ companion,
similar to so-called ``redback'' systems \citep[where outflows from
$\ga0.15$\,\msun\ non-degenerate companions can cause irregular
radio eclipses of the pulsar; e.g.,][]{dpm+01}.  Despite many
observations at essentially all orbital phases, no radio eclipses
have been detected.  This is therefore unlikely to be a redback.
More likely the companion is a white dwarf. If its distance is
$\approx 1.0$\,kpc, as inferred from the DM, this could be a good
target for optical studies; it may be a system similar to PSR~J1012+5307,
an MSP in a 14\,hr orbit with a spectroscopically identified
0.16\,\msun\ white dwarf \citep{kbk96}, or to PSR~J1738+0333
\citep{avk+12}.

\psrj\ is in a 3\,hr orbit with a $>0.021$\,\msun\ companion.  These
parameters suggest a ``black widow'' interacting system \citep[sub-day
binaries with degenerate $\la0.05$\,\msun\ companions; e.g.,][]{fst88}.
Most such systems show radio eclipses near superior conjunction,
but so far we have not detected any (when folded using the known
orbital parameters, the pulsar is detected in all six data sets
listed in Table~\ref{tab:survey}, including two observations of
superior conjunction).  This could be due simply to geometry: if
the actual companion mass is slightly larger than the minimum value
inferred from the mass function, e.g., if $m_2 \ga 0.025$\,\msun,
the orbital inclination angle $i \la 60\arcdeg$ and we could be
viewing the system relatively face-on.  In any case, with a DM-derived
distance of 0.9\,kpc, this may also be an interesting optical target.

It seems curious that among the 10 MSPs discovered in this Parkes
survey only one is in an interacting binary system (either black
widow or redback), when about 50\% of the 67 MSPs so far discovered
in searches of unidentified LAT sources worldwide are in such systems
\citep[see][]{rob13}.  For instance, in a recent survey at Arecibo,
five of six MSPs discovered are either black widows or redbacks
(H.~T.\ Cromartie et al.\ 2015, in preparation). However, the Arecibo
searches have integration times of $T = 15$\,min, which are far
more suitable for the discovery of few-hour binaries than the 1\,hr
Parkes integrations: the maximum accelerations probed by our searches
scale as $T^{-2}$.  At the GBT, where half of these 67 MSPs were
discovered, typical integration times are 30--45\,min \citep[e.g.,][P.\
Bangale et al.\ 2015, in preparation]{rrc+11}.  In any case, the
pulsars that we did not detect unbiasedly in every one of our Parkes
survey observations are the two sub-day binaries (Table~\ref{tab:survey}),
at least in part due to large and rapidly changing orbital acceleration
(cf.\ Figure~\ref{fig:J1946}).

\subsection{Six New Gamma-ray Millisecond Pulsars} \label{sec:sixgamma}

Radio timing observations since 2009 have yielded rotational
ephemerides for 43 of the 67 MSPs discovered in LAT-guided searches.
Of these, 39 are now confirmed as gamma-ray MSPs \citep[the other
four are unrelated to the LAT sources; e.g.,][]{kjr+11}, including
the six discovered in our Parkes survey for which we already have
timing solutions. In this paper we have for the first time presented
the ephemerides and polarimetry for five of these MSPs, and gamma-ray
properties for \psri. In some cases these results add to our
understanding of the pulsars summarized in the second LAT catalog
of gamma-ray pulsars \citep[2PC;][]{aaa+13}.

For \psri, the radio peak leads the gamma-ray profile by $\delta
\approx 0.5\,P$. Other pulsars with such values of $\delta$ have
gamma-ray profiles composed of only one peak (2PC), which is also
the case for \psri\ (Figure~\ref{fig:J1903}). This is one of the
few MSPs for which spectral curvature (deviation from a power-law
spectrum) is not apparent even with 4\,yr of LAT data
(Table~\ref{tab:survey2}).  Our spectral fits to 6.5\,yr of data
yield the largest cut-off energy of any pulsar, even if with large
uncertainty: $E_c = (7.8\pm4.0)$\,GeV (Table~\ref{tab:J1903}; see
also Figure~\ref{fig:3fglspectra}f).  The MSPs with the next largest
values of $E_c$ are also from our Parkes sample: PSRs~J1747$-$4036
and J1514$-$4946 (2PC).

With our proper motion measurement for \psra\ (Table~\ref{tab:parms}),
its intrinsic $\dot E$ is 1.6 times smaller than previously thought.
Its computed gamma-ray efficiency, already high in 2PC (30\%), now
increases to 50\%. As usual this assumes a geometry-dependent beaming
correction of $f_\Omega = 1$ (see 2PC for definition), i.e., assuming
isotropic emission. Perhaps for \psra, $f_\Omega \ll 1$.  Or maybe
its DM-derived distance of 0.9\,kpc is an overestimate \citep[its
nominal transverse velocity is the largest of those listed in
Tables~\ref{tab:parms} and \ref{tab:J1903}, although it is not
unusually large for an MSP; e.g.,][]{gsf+11}.

\psrc\ is notable in several respects. Spectral curvature is not
apparent in the 3FGL catalog (see Table~\ref{tab:survey2}).  Its
$\dot E$ is large (fourth highest among the 2PC MSP sample).  Its
nominal distance (Table~\ref{tab:parms}) is also among the largest
in the 2PC MSP sample, and it is a relatively faint gamma-ray source
($10\,\sigma$ in 3FGL; Table~\ref{tab:survey2}). If it had a typical
$\dot E$ at that distance presumably it would not be a detectable
LAT pulsar. This large $\dot E$ ultimately is due to its very short
spin period (third shortest among disk MSPs). While \psrc\ may be
somewhat extreme in this regard, it does appear that LAT-detected
MSPs are a smaller-$P$, higher-$\dot E$ population than radio-selected
MSPs \citep{rap+12}. In our Parkes sample, PSRs~J1902$-$5105 and
J0955$-$6150 (the latter apparently quite distant) are also
short-period MSPs, with $P<2$\,ms.

\psrd\ is one of only six MSPs known with phase-aligned gamma-ray
and radio light curves (2PC).  This points to emission that is
either extended, and caustic in nature, or originates near the
neutron star surface \citep{vjh12}.  In order to constrain their
emission and viewing geometry, \citet{jvh+14} have jointly modeled
the gamma-ray and radio profiles of 40 LAT-detected MSPs. They fit
gamma-ray light curves using standard outer magnetosphere gap (OG)
and two-pole caustic (TPC) geometric models assuming a vacuum
retarded dipole magnetic field. For \psrd, \citet{jvh+14} consider
altitude-limited (al) versions of the standard models as well as a
low-altitude slot gap model (laSG). They find that all three models
match the gamma-ray and radio profiles fairly well.  However, the
complete lack of polarization that we observe (Figure~\ref{fig:pol}d)
strongly favors the alTPC and alOG models, where high-altitude
caustics are predicted to have a strong depolarizing effect
\citep{dhr04}, over the laSG model, where one expects non-zero
polarization from near-surface emission.

For \psra, the high level of observed linear polarization and modest
P.A.\ swing (Figure~\ref{fig:pol}a and Section~\ref{sec:pol}) suggest
we are viewing the edge of a cone beam at relatively low altitude.
For \psrb, the very high level of linear polarization suggests that
at least some components originate at relatively low altitude (the
small peak with low polarization, which might also originate from
the opposite pole, could be a high-altitude caustic;
Figure~\ref{fig:pol}b).  \psrc\ has a very unusual polarization
pattern (Figure~\ref{fig:pol}c), which does not fit either standard
radio core/cone or high-altitude caustic emission models.  For these
MSPs, \citet{jvh+14} find models that plausibly match the gamma-ray
light curves, but the radio profiles are not well reproduced.

\subsection{Survey Statistics} \label{sec:stats}

Our survey of unidentified LAT sources had a success rate of 20\%
(11 MSPs detected unbiasedly in 56 sources searched;
Section~\ref{sec:searches}), despite significant selection effects
(Section~\ref{sec:sensitivity}).  This is very encouraging --- a
substantial number of unidentified LAT sources contain previously
unknown MSPs that can be detected at Parkes, and our source selection
criteria have allowed us to target them with satisfactory efficiency.
However, comparing the initial segment of the survey
(Section~\ref{sec:1fgl}) with the latter portion (Section~\ref{sec:2fgl})
reveals a drop in efficiency: we had to search 3.5 times as many
sources in 2012, for twice as long in the aggregate, to discover
as many pulsars as earlier on.

Since our observations, another six pulsars have been discovered
in these very same sources: four via direct pulsation searches of
the gamma-ray photons, one of which was subsequently detected as a
very faint radio source with the GBT (the others might be significantly
affected by scintillation, or perhaps they may yet prove to be
gamma-ray-only pulsars); one at the Giant Metrewave Radio Telescope
(GMRT) at a lower frequency than at Parkes (perhaps the Parkes
non-detections reflect a steep radio spectrum); and one that is a
transient radio source --- not active when we did our Parkes searches,
but now detectable there (see notes l--o in Table~\ref{tab:survey2}).
Considering these additional detections, a full 10 of the original
14 targets of our survey are now known to contain pulsars! This
reflects superb source selection criteria for those targets. Why
has the success rate decreased since then? The original (1FGL-based)
sources were brighter in gamma rays, allowing for relatively
unambiguous determination of spectral characteristics, on which we
based our target selection. There might be secondary effects related
to this: e.g., the brighter 1FGL-based targets might be nearer to
the Earth on average, and any associated radio pulsars could then
also be brighter on average, although we do not see such an effect
among our small sample of {\em detected\/} MSPs.

\subsection{Spectral Characteristics of LAT Pulsars}
\label{sec:psrclass}

\begin{figure*}
\begin{center}
\includegraphics[scale=0.62]{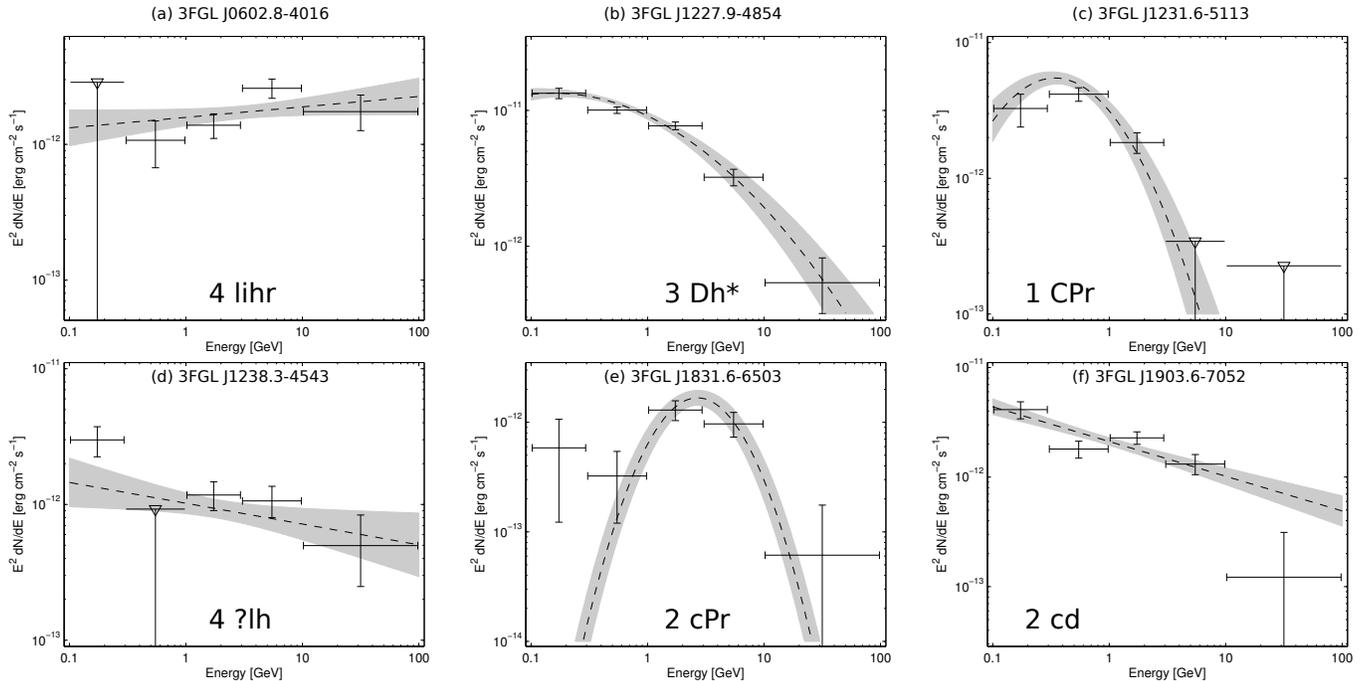}
\end{center}
\caption{\label{fig:3fglspectra}
Examples of spectral classification of unidentified \fermi-LAT
sources with a view towards identifying pulsar candidates.  The
panels show the spectral energy distribution (SED) from the 3FGL
analysis \citep{aaa+15}, with the dashed lines/gray bands indicating
the best-fit model SED (either a power law or a curved log-parabolic
spectrum) and its uncertainties.  Discrepancies between the spectral
measurements and the model indicate that the fit may be unreliable.
Our classifications (1--5) and qualitative spectral characteristics
that underlie them are described in Section~\ref{sec:unidclass}.
The full list of 56 classified sources searched in our Parkes radio
survey is given in Table~\ref{tab:survey2}.  \textbf{(a)} An example
of a poor pulsar candidate, consistent with a power law (l), with
both high-energy emission (h) and an AGN association; \textbf{(b)}
an example of an atypical pulsar spectrum, now associated with the
state-changing MSP J1227$-$4853; \textbf{(c)} an excellent candidate
with a strong cutoff (C), a spectrum slightly more peaked than that
of the typical pulsar (P), and rising at $<1$\,GeV (r); \textbf{(d)}
a poor candidate spectrum with both high-energy emission (h) and
unreliable spectral points (?); \textbf{(e)} a good candidate with
an unusually peaked spectrum (P); \textbf{(f)} a decent candidate
with evidence for a cutoff (c) but an unusually steep low-energy
spectrum, monotonically decreasing (d); now known to be \psri\ (this
work; e.g., see Table~\ref{tab:J1903} and top panel of
Figure~\ref{fig:J1903}).  }
\end{figure*}

Now that the third catalog of LAT sources is in hand, based on
substantially more data and a much better understanding of the
instrument and background than was available earlier, we can usefully
revisit our 56 sources and the MSPs found amid them.  In
Table~\ref{tab:survey2} we have summarized some properties of these
sources as obtained from the 3FGL catalog.  Given that gamma-ray
pulsars typically have exponentially cut-off power-law spectra
\citep{aaa+13}, by contrast to power-law spectra for AGN, it is not
surprising that the spectral curvature gleaned from 3FGL correlates
well with gamma-ray pulsars: of the 12 pulsars with known gamma-ray
pulsations listed in Table~\ref{tab:survey2} (the eight classified
as ``PSR,'' as well as J1035.7$-$6720, J1227.9$-$4854, J1624.2$-$4041,
and J1744.1$-$7619), 10 display significant curvature according to
3FGL ($\mbox{Curve}>4\,\sigma$; Table~\ref{tab:survey2}).  However,
two other established gamma-ray MSPs (PSRs~J1747$-$4036 and
J1903$-$7051, reported here) show no curvature, according to 3FGL
($\mbox{Curve}<3\,\sigma$).  Now that 170 gamma-ray pulsars are
known, it is to be expected that some may depart from the norm. In
addition, ``Curve'' in 3FGL tests against a log-parabolic spectral
model, which may not be a good proxy for testing against exponentially
cut-off power laws, especially when the source is not very bright
or the background is problematic (as it happens, these two MSPs are
the faintest in gamma rays of the 12 under consideration; see ``Sig''
in Table~\ref{tab:survey2}).  Thus, in considering whether a
particular 3FGL source is a good pulsar candidate, we should do
more than simply look for significant cataloged curvature.

\subsection{Spectral Classification of Unidentified LAT Sources}
\label{sec:unidclass}

Flux information is available from the 3FGL pipeline processing in
five energy bins (two within 0.1--1\,GeV, two within 1--10\,GeV,
and one for 10--100\,GeV).  After familiarizing ourselves with how
known gamma-ray pulsars (and control sources) appear in five-bin
spectra by visual inspection of many 3FGL plots, we have classified
the 56 target sources of our Parkes survey according to a heuristic
ranking scheme that we describe next.

We qualitatively assess the likelihood of a source being a pulsar.
According to this scheme, a classification of ``1'' denotes near
certainty of being a pulsar; ``2'' is less conclusive but quite
plausible; ``3'' is a poorer pulsar candidate; ``4'' is very likely
not a pulsar; ``5'' is not a pulsar.

These classifications are based on combinations of the following
spectral characteristics (both are listed for each source in
Table~\ref{tab:survey2}): strong energy cutoff (c); parabolic (peak
at center energies, p); flat or rising spectrum at $<1$\,GeV (r)
--- c, p, r are ``positive'' features leading to higher likelihood
of a source being a pulsar (for examples, see Figure~\ref{fig:3fglspectra}c
and e).  Power law (l); variability (v); monotonically decreasing
(d) or increasing (i); excess high energy emission (flat or rising
spectrum; h) --- l, v, d, i, h are ``negative'' features, not
ordinarily associated with pulsars (see Figure~\ref{fig:3fglspectra}a
and d).  Capitalization means that the features are more certain,
except for p/P which indicates the sharpness of the parabola.  A
``?'' indicates poor spectral quality, increasing uncertainty in
classification.  An asterisk represents an odd spectrum with
``banana'' shape to high energy (see Figure~\ref{fig:3fglspectra}b).
In Figure~\ref{fig:3fglspectra} we show 3FGL spectral plots for six
of our Parkes targets that illustrate the features described above.

Sixteen of our 56 sources are classified as 5 or 4 (or 3 with a
blazar association), and we regard them as no longer viable pulsar
candidates.

Of the 12 known gamma-ray pulsars in Table~\ref{tab:survey2}, nine
are classified as 1 and two others (already noted in
Section~\ref{sec:psrclass} as not curved in 3FGL) as 2. The one
confirmed gamma-ray MSP classified as a 3, PSR~J1227$-$4853, is a
spectral outlier (Figure~\ref{fig:3fglspectra}b). In a rare
state-changing binary system, it is borderline variable within 3FGL,
and recently displayed significant variability \citep[see][]{jrr+15}.
This and its sister system J1023+0038 \citep{sah+14}, as well as
the young pulsar J2021+4026 \citep{abb+13}, are counterexamples to
the usual assumption that pulsars are steady gamma-ray emitters.
Of the other five MSPs known in Table~\ref{tab:survey2} (none of
which has yet a reported long-term timing solution or detected
gamma-ray pulsations), one is classified as 1, one as 2, and three
as 3. Based on prior statistics, we expect that most, perhaps all,
of these five MSPs will eventually be established as gamma-ray
pulsars.  That most have a relatively poor classification in our
scheme likely reflects their faintness (three, all discovered in
our survey, have 3FGL significance of 6.6--8.8$\,\sigma$), or
location close to the Galactic plane (PSR~J1536$-$4948 is at
$b=4.8\arcdeg$).

\subsection{Further Searches of LAT Sources} \label{sec:future}

Following from the discussion in Section~\ref{sec:unidclass}, while
we certainly recommend additional searches first of the ``1'' sources
that remain without coincident pulsars (of which there are seven
in Table~\ref{tab:survey2}), followed by the ``2'' sources (nine
in Table~\ref{tab:survey2}), the ``3'' sources (seven in
Table~\ref{tab:survey2}) are also reasonable candidates for further
searches\footnote{One of the ``2'' sources, 3FGL~J1417.5$-$4402,
has recently been reported to be a binary system currently with an
accretion disk \citep{scc+15}. }. Among these 23 sources, two are
coincident with globular clusters that currently have no known
pulsars, and a population of MSPs in those clusters is a plausible
origin for the gamma-ray emission. We expect further MSP discoveries
among these 23 targets: while the largest offset between our optimized
search and the 3FGL positions is 3\arcmin, which is not very large
compared to the $7\farcm2$ half-width at half-maximum Parkes beam,
many promising targets have not been searched enough to counter the
selection effects presented in Section~\ref{sec:sensitivity} ---
e.g., four of the seven ``1'' sources have been searched at Parkes
only once or twice at locations not far from their 3FGL positions
(see Table~\ref{tab:survey2}). In addition, in the near future we
intend to re-analyze our existing data sets with ``jerk searches,''
i.e., accounting for changing accelerations that are relevant for
short orbital periods (cf.\ Figure~\ref{fig:J1946}).

After all such searches are done, it remains a possibility that
through a combination of faintness and extreme orbital and spin
parameters, some of those promising LAT sources could still harbor
undetected radio MSPs beamed towards the Earth.  It is also possible
that some of these sources may be MSPs that are detectable only via
their gamma-ray emission, although this fraction is known to be
small \citep[e.g.,][]{rom12}.

\acknowledgements

We are grateful to the wonderful staff at the Parkes telescope for
their role in making it such a fine research instrument. We thank
in particular the friends of the analog filterbank system, used to
discover more than 1000 pulsars, who staved off its demise long
enough to allow the completion of this work.  We thank Willem van
Straten and Paul Demorest for their help with PSRCHIVE.  The Parkes
Observatory is part of the Australia Telescope, which is funded by
the Commonwealth of Australia for operation as a National Facility
managed by CSIRO.

The National Radio Astronomy Observatory is a facility of the
National Science Foundation operated under cooperative agreement
by Associated Universities, Inc.

The \fermi\ LAT Collaboration acknowledges generous ongoing support
from a number of agencies and institutes that have supported both
the development and the operation of the LAT as well as scientific
data analysis.  These include the National Aeronautics and Space
Administration (NASA) and the Department of Energy in the United
States, the Commissariat \`a l'Energie Atomique and the Centre
National de la Recherche Scientifique / Institut National de Physique
Nucl\'eaire et de Physique des Particules in France, the Agenzia
Spaziale Italiana and the Istituto Nazionale di Fisica Nucleare in
Italy, the Ministry of Education, Culture, Sports, Science and
Technology (MEXT), High Energy Accelerator Research Organization
(KEK) and Japan Aerospace Exploration Agency (JAXA) in Japan, and
the K.~A.~Wallenberg Foundation, the Swedish Research Council and
the Swedish National Space Board in Sweden.

{\em Facilities:}  \facility{Parkes (PMDAQ, PDFB)}, \facility{GBT
(GUPPI)}, \facility{Fermi (LAT)}, \facility{CXO (ACIS-S)},
\facility{Swift (XRT)}

\end{document}